\documentclass[12pt,a4paper,notitlepage]{article}

\usepackage{graphicx}
\usepackage{caption}
\usepackage{subcaption} 
\usepackage{authblk}
\usepackage{multirow}
\usepackage{amsmath} 
\usepackage[numbers, sort&compress]{natbib} 
\usepackage{hyperref}
\usepackage{doi}

\newcommand{\scifi}{SciFi}

\newcommand\parbar[1]{\overset{\textbf{\fontsize{2pt}{2pt}\selectfont(---)}}{#1}}
\newcommand{\spadwithlens}{SwissSPAD2-$\mu$\textrm{lens}}
\newcommand{\spadwithoutlens}{SwissSPAD2}
\newcommand{\SrY}{$^{90}$Sr}

\title{Demonstration of particle tracking with scintillating fibres read out by a SPAD array sensor and application as a neutrino active target}

\author[1]{Matthew Franks}
\author[1]{Till Dieminger}
\author[2]{Kodai Kaneyasu}
\author[1]{Davide Sgalaberna\thanks{Corresponding author. Email: davide.sgalaberna@cern.ch}}
\author[2]{Claudio Bruschini}
\author[2]{Edoardo Charbon}
\author[1]{Umut Kose}
\author[1]{Botao Li}
\author[2]{Paul Mos}
\author[2]{Michael Wayne}
\author[1]{Tim Weber}
\author[1]{Jialin Wu}

\affil[1]{Institute for Particle Physics and Astrophysics (IPA), ETH Z\"urich, Ramistrasse, Z\"urich, 8093, Switzerland} 
\affil[2]{Advanced Quantum Architecture Lab (AQUA), EPFL, Rue de la Maladi\`ere, Neuch\^atel, 2000, Switzerland} 

\begin{document}
\maketitle
\begin{abstract}
Scintillating fibre detectors 
combine sub-mm resolution particle tracking, precise measurements of the particle stopping power and sub-ns time resolution. Typically, fibres are read out with silicon photomultipliers (SiPM). Hence, if fibres with a few hundred $\mu$m diameter are used, either they are grouped together and coupled with a single SiPM, losing spatial resolution, or a very large number of electronic channels is required.
In this article we propose and provide a first demonstration of a novel configuration which allows each individual scintillating fibre to be read out regardless of the size of its diameter,
by imaging them with Single-Photon Avalanche Diode (SPAD) array sensors.
Differently from SiPMs, SPAD array sensors provide single-photon detection with single-pixel
spatial resolution.
In addition, O(us) or faster coincidence of detected photons allows to obtain noise-free images.
Such a concept can be particularly advantageous if adopted as a neutrino active target, where scintillating fibres alternated along orthogonal directions can provide isotropic, high-resolution tracking in a dense material and reconstruct the kinematics of low-momentum protons (down to 150 MeV/c), crucial for an accurate characterisation of the neutrino-nucleus cross section. 
In this work the tracking capabilities of a bundle of scintillating fibres coupled to SwissSPAD2  
is demonstrated. 
The impact of such detector configuration in GeV-neutrino experiments is studied with simulations and reported.
Finally, future plans, including the development of a new SPAD array sensor optimised for neutrino detection, are discussed. 
\end{abstract}

\small{Keywords: Particle Detectors, SPAD array sensor, Scintillating fibres, Neutrino}
\maketitle

\section{Introduction}
\label{sec:introduction}
Long-baseline neutrino oscillation (LBL) experiments are searching for the violation of the charge-parity (CP) symmetry in the leptonic sector and aim to define the neutrino mass ordering thanks to high-intensity neutrino beams
and much larger masses (from several tenths to a few hundred kilotonnes) at the far detector (FD), located hundreds kilometers away from the neutrino target production, where the oscillation probability is near maximum \cite{HK-LOI,Abi:2020qib,Abe:2019vii,NOvA:2022see}. 
The substantial reduction of the statistical uncertainties will enhance the sensitivity to the neutrino oscillation parameters.
However, in order to achieve the target sensitivity to the CP-violating phase
a drastic reduction of the systematic uncertainties is necessary. 
The key is the so-called Near Detector (ND) that, with a mass that typically ranges from a few hundred kilograms to several tonnes, is placed only a few hundred meters from the neutrino production target i.e., before neutrinos can oscillate.
Its goal is the precise detection of neutrino interactions to measure the neutrino flux and the neutrino-nucleus cross section.
The latter is largely affected by the so-called nuclear effects,
such as Fermi motion, binding energy, nucleon-nucleon correlation also called 2 particle - 2 holes (2p2h) 
and final-state interactions,
that can introduce a significant smearing and a bias in the reconstructed neutrino energy and, if not properly modelled, can lead to a wrong measurement of the neutrino oscillation parameters \cite{Ankowski:2017yvm}. New kinematical analysis methods of the neutrino interaction final state have been developed in order to distinguish the various nuclear processes and improve the modelling of the nuclear effects \cite{Lu:2015tcr,Lu:2015hea}. 
However, their full exploitation requires beyond state-of-the-art
detectors
to identify and reconstruct the kinematics of all the hadrons in the neutrino interaction final state, typically protons, neutrons, pions and, possibly, even bound nuclei \cite{Ershova:2022jah}.
This will be key in reducing the systematic uncertainties at the future neutrino LBL experiments.

Current and future neutrino detectors, typically based on liquid argon \cite{Acciarri:2016smi,ICARUS:2023gpo,SBND:2020scp,Abi:2020qib} or organic plastic scintillator
\cite{Amaudruz:2012esa,SoLid:2018jas} 
have improved the state of the art, by combining particle tracking over the full solid angle with calorimetry, and reaching proton detection thresholds down to $\sim300~\text{MeV/c}$. 
Moreover, recently developed 3D granular plastic scintillator detectors \cite{Sgalaberna:2017khy},
will aim at high-efficiency neutron detection with energy measurement using the time-of-flight technique \cite{Munteanu:2019llq,ND280upgrade-tdr,Gwon:2022bix,Agarwal:2022kiv}.
However, none of these detector technologies has a granularity fine enough to detect low-energy protons that, down to momenta below $200~\text{MeV/c}$ leave tracks only a few-mm in length. 
For such a purpose, High-Pressure Time Projection Chambers (HpTPC) have been proposed, 
however, they are still low-density and require very large magnetised volumes, hence high costs, to reach a O(1~tonne) mass.

The solution can be found in a scintillating fibre detector (\scifi) used as neutrino active target that, thanks to a sub-$100~\mu\text{m}$ spatial resolution \cite{Joram:2015ymp,Papa:2023uqv},
can easily reconstruct sub-mm proton tracks. 
\scifi~detectors have been used in neutrino experiments as tracking detectors but not as neutrino active target \cite{Eskut:2007rn,SNDLHC:2023pun}.
Although similar designs have been proposed \cite{Annis:1997es}, their read out based on CCD shows a relatively high contamination from noise.

Typically read out with silicon photomultipliers (SiPMs), one limitation is that to obtain massive detectors necessary to collect enough neutrino interactions whilst preserving an excellent spatial resolution, 
a prohibitive number of electronic readout channels is needed, with consequent large costs. 
Hence, a compromise is achieved by coupling a single SiPM 
(e.g. $1.25\times0.25\,\text{mm}^2$) with a bundle of thinner fibres ($0.25~\text{mm}$ diameter) sacrificing the spatial resolution on one spatial coordinate \cite{Joram:2015ymp}.

Recently, a new type of photosensor technology has seen tremendous improvements. It consists of Single-Photon Avalanche Diode (SPAD) array sensors. 
Whilst SiPMs simply count the total number of scintillation photons detected within its own active area with O(100~ps) time resolution, SPAD arrays can provide both the position and the time of each single detected photon with a spatial resolution driven by the size of the single pixel pitch. 
Hence, each SPAD pixel can be considered as an independent readout channel.
This is made possible by placing the circuitry for digitisation of single-SPAD signal on the chip near the pixels, at the expense of a slightly lower fill factor.
Depending on the application that includes LIDAR \cite{DBLP:ZhangLAPWC19},
Non-Line-of-Sight Imaging \cite{spad-nonlineofsightimaging}, 
and Raman Spectroscopy \cite{spad-biophotonics},
different SPAD arrays and SPAD test structures reported in literature demonstrate a pitch down to just above $2~\mu\text{m}$ \cite{MorimotoOPEX:20,Shimada:22}, time gate based readout, as well as
multi-pixel time-stamping using time-to-digital converters (TDCs) with $\sim 50~\text{ps}$ timestamp resolution \cite{DBLP:ZhangLAPWC19}.
In \cite{Fischer:2022sfe}, SPAD array sensors have been proposed to readout \scifi. However, particle tracking has not been demonstrated with prototypes yet.

In this article, we report the demonstration of particle tracking with SwissSPAD2~\cite{SwissSPAD2Ulku2019} and propose a \scifi~detector read out with SPAD array sensors as a 
cost-effective solution for high-resolution neutrino active target detectors.
The demonstration of particle tracking is obtained by exposing a bundle of scintillating fibres to \SrY~electrons and reading out with SwissSPAD2.
In the last part of the article, the simulation of a SPAD array based \scifi~neutrino detector is studied both considering the SwissSPAD2 response parameters, as well as an improved design that the authors are currently developing.
Such configuration would allow a reduction of the number of readout channels,
which would scale with the detector surface rather than the number of scintillating fibres,
and, at the same time, preserve the finest spatial resolution possible
in a massive fully-active scintillator volume so that even low-momentum 
protons can be tracked. 
Finally, plans for the future R\&D are discussed. 

\section{Demonstration of readout of scintillating fibres with SwissSPAD2}
\label{sec:test-scifi-spad}

A demonstrator consisting of \scifi~read out with the SwissSPAD2 sensor~\cite{SwissSPAD2Ulku2019} has been built and studied. SwissSPAD2 is a $512\times 512$ pixel SPAD array sensor with
a pixel pitch of $16.38~\mu\text{m}$, resulting in total surface of $8.4\times8.4~\text{mm}^2$, of which $10.5\%$ is active.
The Photon Detection Probability (PDP) of these pixels peaks at a wavelength of around 500~nm with values of $55\%$, using a $7~\text{V}$ excess bias $V_{\text{ex}}$.
During operation, the exposure time of one frame $t_{\text{gate}}$ can be tuned from $10~\text{ns}$ to $10~\mu\text{s}$, and data from the whole array is read out in about $10~\mu\text{s}$.
\par
\begin{figure*}
    \centering
    \begin{subfigure}{0.44\textwidth}
        \centering
        \includegraphics[width=\linewidth]{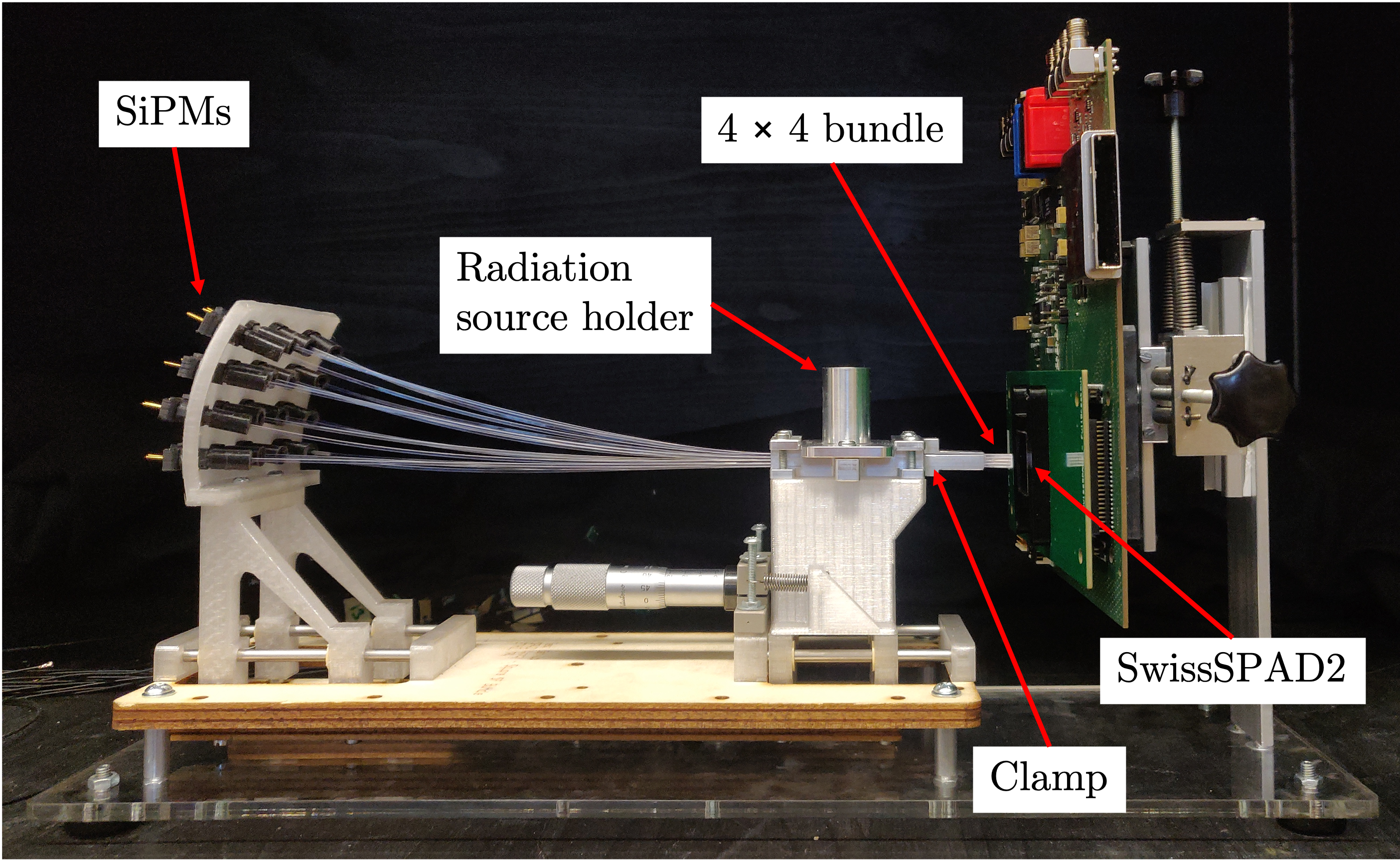}
        \caption{}
        \label{fig:SciFi_setup}
    \end{subfigure}%
    \hspace{0.001\textwidth}
    \begin{subfigure}{0.27\textwidth}
        \centering
        \includegraphics[width=\linewidth, clip, trim=5cm 5cm 5cm 5cm]{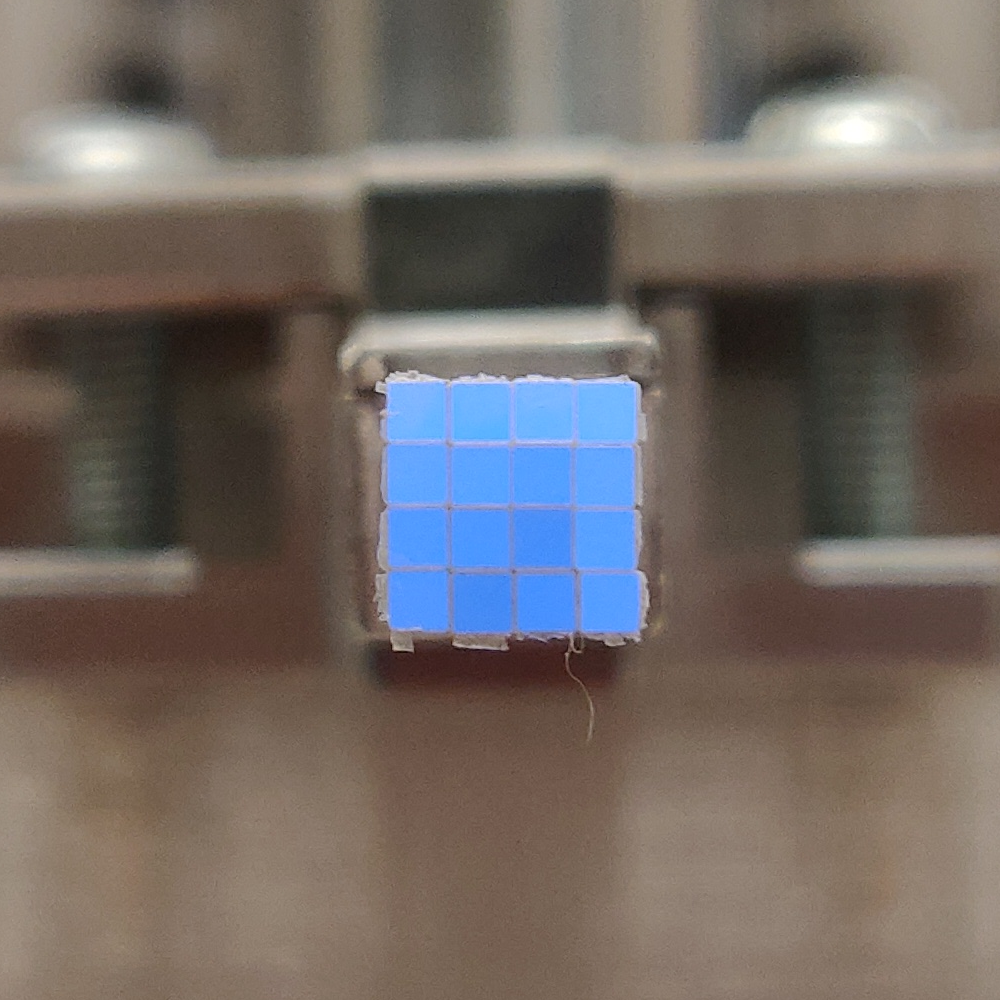}
        \caption{}
        \label{fig:SciFi_bundle}
    \end{subfigure}%
    \hspace{0.001\textwidth}
    \begin{subfigure}{0.27\textwidth}
        \centering
        \includegraphics[width=\linewidth, clip, trim=10cm 10cm 10cm 10cm]{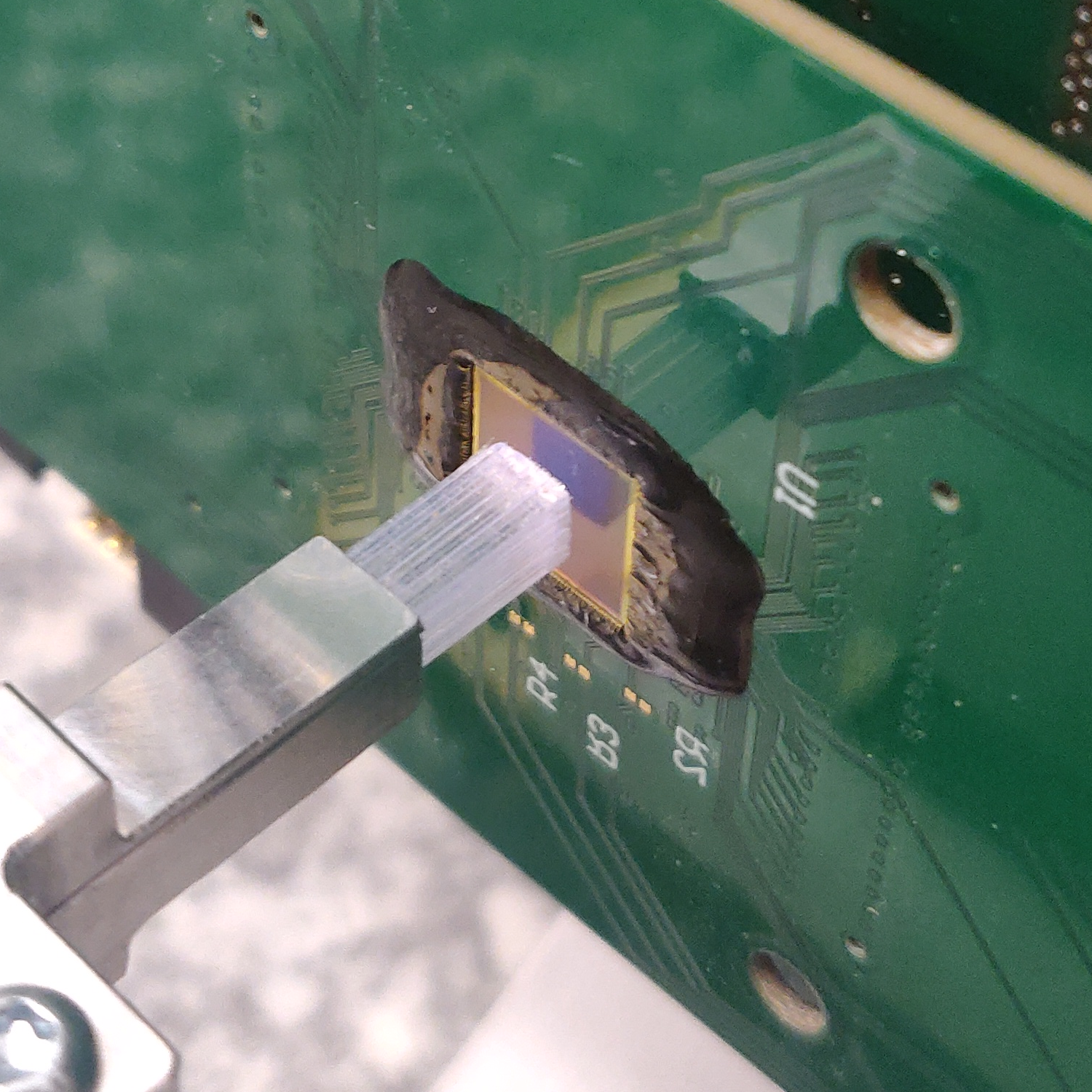}
        \caption{}
        \label{fig:SciFi_setup_zoom}
    \end{subfigure}
    \caption{
    (a) SciFi setup used for experiments. Scintillating fibres are independently coupled to SiPM on their left side end, while on the opposite side they are bunlded together and coupled to a single SwissSPAD2 sensor.
    (b) Close-up photograph of the face of $4\times4$ scintillating fibre bundle.
    (c) Coupling of the scintillating fibre bundle with SwissSPAD2. Black glob top was used to protect the wire bonding around the active area.
    }
    \label{fig:scifi-prototype}
\end{figure*}

 A first prototype of the proposed SciFi detector concept, shown in Fig. \ref{fig:SciFi_setup}, was constructed using sixteen $25~\text{cm}$ long scintillating fibres bundled in a $4\times4$ grid (see Fig. \ref{fig:SciFi_bundle}), using a machined aluminum clamp,
with around $1~\text{cm}$ of fibre protruding from the end of the clamp to couple to SwissSPAD2.

The main goal was the detection of individual tracks created by minimum ionising particles crossing a scintillating fibre detector read out, for the first time, by a SPAD array sensor. 
Its performance was compared to a readout based on SiPMs, typical for this type of detector which were coupled to the opposite end of the fibres. 
Kuraray SCSF-78F $1 \times 1~\text{mm}^2$, square, single-cladding, plastic scintillating fibres emitting in the blue around 450~nm \cite{kuraray-catalogue-ps}
were used.
The surface of these fibres was polished with diamond lapping sheet down to $3~\mu\text{m}$ grit to ensure a smooth finish and minimise internal reflections to increase the light yield.

On top of the aluminum clamp a radiation source holder was fastened, allowing the mounting of a blue flashlight (450 nm nominal wavelength), which was used to align the SPAD array with the SciFi bundle, tune the respective distance, and identify the region of pixels detecting the visible photons from each different scintillating fibre.
A 3D printed black plastic collimator with a 100 $\mu$m hole allowed a more precise exposure onto a single column of fibres, with a reduced light intensity to avoid saturation on the SPAD array.

In a similar manner, the scintillating fibres were also exposed to a \SrY ~radioactive source ($185^{+0}_{-40\%}~\text{kBq}$), placed directly on top of the bundle.
The \SrY~source undergoes a series of two $\beta^-$ decays. The first one results in $^{90}\text{Y}$ with a Q-value of 0.56 MeV, while the second one in $^{90}\text{Zr}$ with a Q-value of 2.28 MeV. From Geant4 simulations one can see that electrons with kinetic energies below 1 MeV have a typical range of less than 4~mm in plastic with almost half of the energy loss in the last mm. Instead, more energetic electrons can range up to 10~mm with a more uniform energy deposition.
Given the presence of a $0.1~\text{mm}$ stainless steel foil directly under the radioactive material inside the source, only electrons with an energy above 250 keV and a maximum energy of about 2~MeV are expected to reach the fibres~\cite{ROY2018134}.
The range in polystyrene is expected to be longer than the 4~mm thick bundle for endpoint electrons. The whole SciFi setup, including the photon readout system (described below), was put on rails, which could be positioned using a micrometer screw for fine adjustments of the fibre end to SPAD array distance.

The fibre bundle was coupled on one side with a single SwissSPAD2 as shown in Fig. \ref{fig:SciFi_setup_zoom}. Although not in contact, the distance between the fibre end and the SwissSPAD2 active area was less than 0.5~mm.
The nominal SwissSPAD2 Photon Detection Efficiency (PDE) with $V_{\text{ex}}$ equal or larger than 5V is higher than 4~\% at 450~nm \cite{SwissSPAD2Ulku2019}, the peak of SCSF-78 fibre emission spectrum ~\cite{kuraray-catalogue-ps}, and up to about 5\% at 500~nm.

A second configuration of SwissSPAD2 was also tested. In order to increase the effective fill factor, each single pixel was coupled to a spherical refractive microlens.
Its goal is to increase the fraction of incident light impinging on the pixel to its active area. The details of the microlens design can be found in \cite{Bruschini:23}.
Based on design simulations, the fill factor can increase up to a factor five \cite{SwissSPAD2Ulku2019}.
However, given the 0.55 nominal numerical aperture of the used scintillating fibres \cite{kuraray-catalogue-ps}, the effective fill factor is not expected to achieve its maximum theoretical value.
Later in the text we will name \spadwithlens~the SwissSPAD2 with a microlens on each pixel and \spadwithoutlens~the one with naked pixels.

To monitor the temperature close to the SPAD array, a bluetooth thermosensor was placed close to the chip and monitored during the data acquisition.

On the opposite end of the SciFi bundle, each individual fibre was polished and coupled with a SiPM, thanks to 3D printed black optical connectors. Hamamatsu ceramic-type Multi-Pixel PhotoCounters (MPPCs) with 50~$\mu$m pitch (S13360-1350CS), for a nominal PDE of 40\%, \cite{hamamatsu:mppc} were used. These where mounted in a custom 3D printed grid as seen on the left in Fig.~\ref{fig:SciFi_setup}. 

The whole setup was placed in a dark cabinet and covered with a black cloth to protect from environmental light.

\subsection{Setup characterisation}
\label{sec:test-characterisation}
Before exposing the SciFi bundle to \SrY~electrons, the setup was fully characterised: the position of each fibre on the SPAD array, noisy pixels were identified, and pixel-to-pixel crosstalk was evaluated.

\subsubsection*{Fibre positions on SPAD array sensor}
First, measurements with a collimated, blue flashlight, positioned above the SciFi bundle were taken to configure and characterise the setup. The same procedure was followed for both \spadwithlens~and \spadwithoutlens.

\begin{figure*}
    \centering
    \begin{subfigure}[b]{0.49\textwidth}
        \centering
        \begin{subfigure}{0.5\textwidth}
            \includegraphics[width=\linewidth]{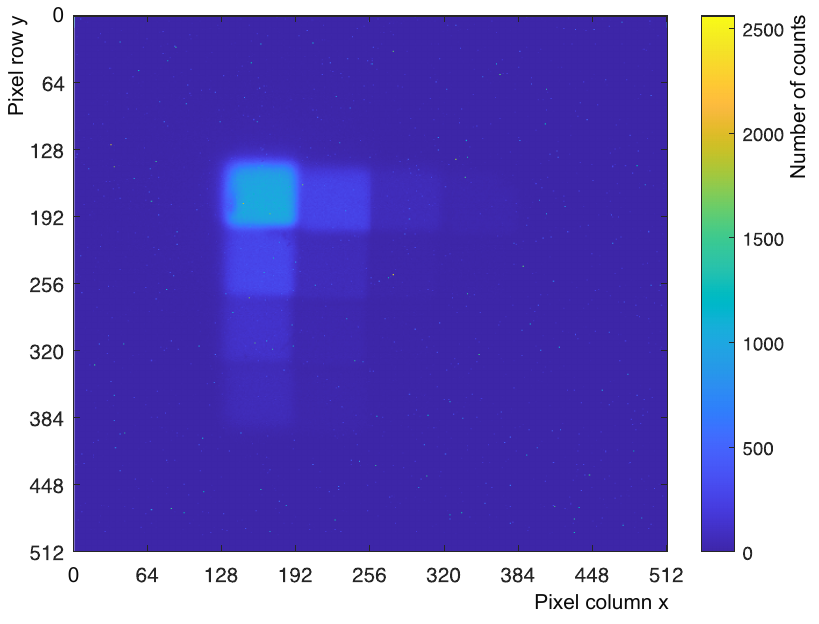}
        \end{subfigure}%
        \begin{subfigure}{0.5\textwidth}
            \includegraphics[width=\linewidth]{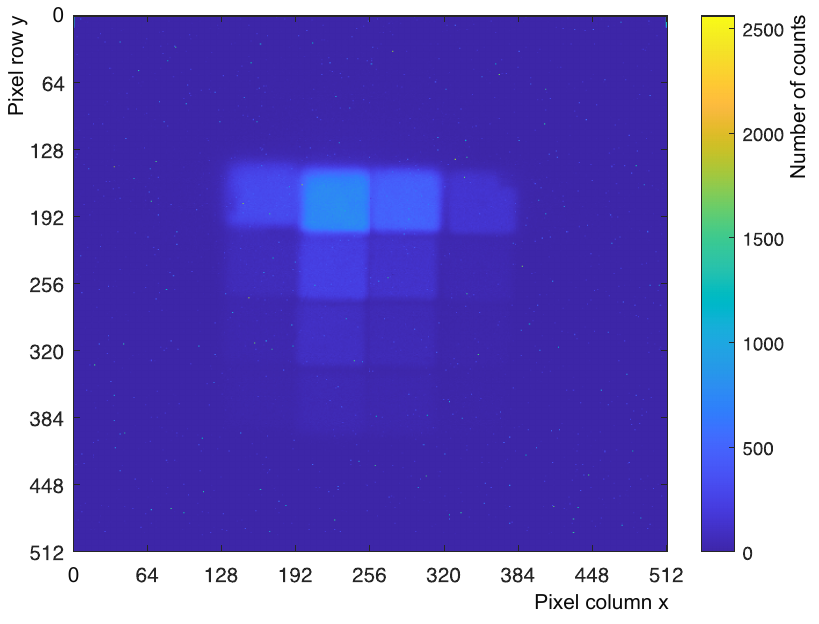}
        \end{subfigure}
        \begin{subfigure}{0.5\textwidth}
            \includegraphics[width=\linewidth]{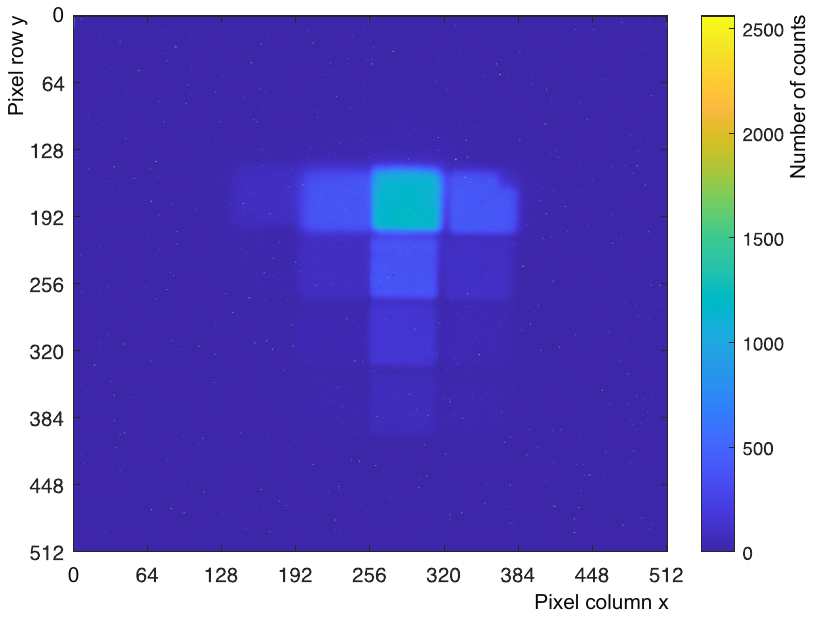}
        \end{subfigure}%
        \begin{subfigure}{0.5\textwidth}
            \includegraphics[width=\linewidth]{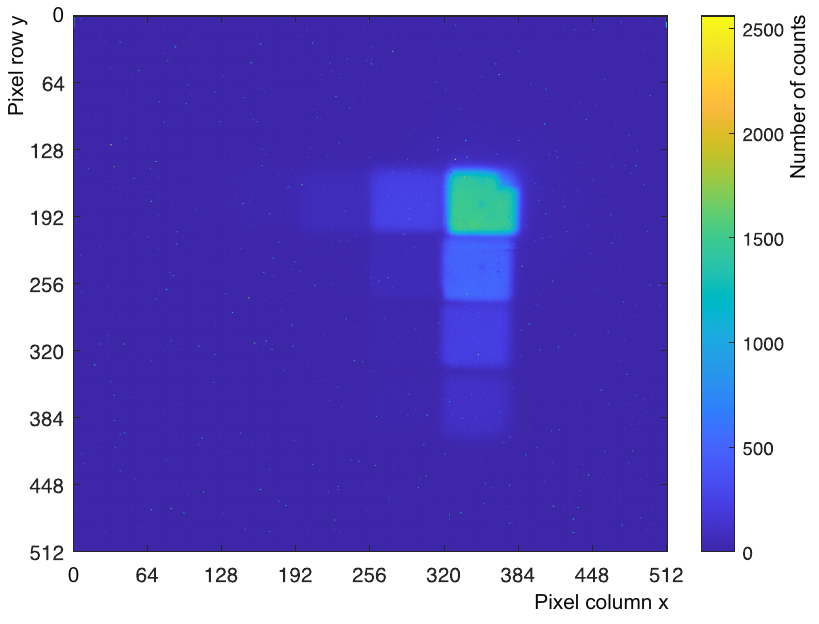}
        \end{subfigure}
        \caption{}
        \label{fig:UV_columns_individual}
    \end{subfigure}%
    \begin{subfigure}[b]{0.49\textwidth}
        \centering
        \includegraphics[width=\linewidth]{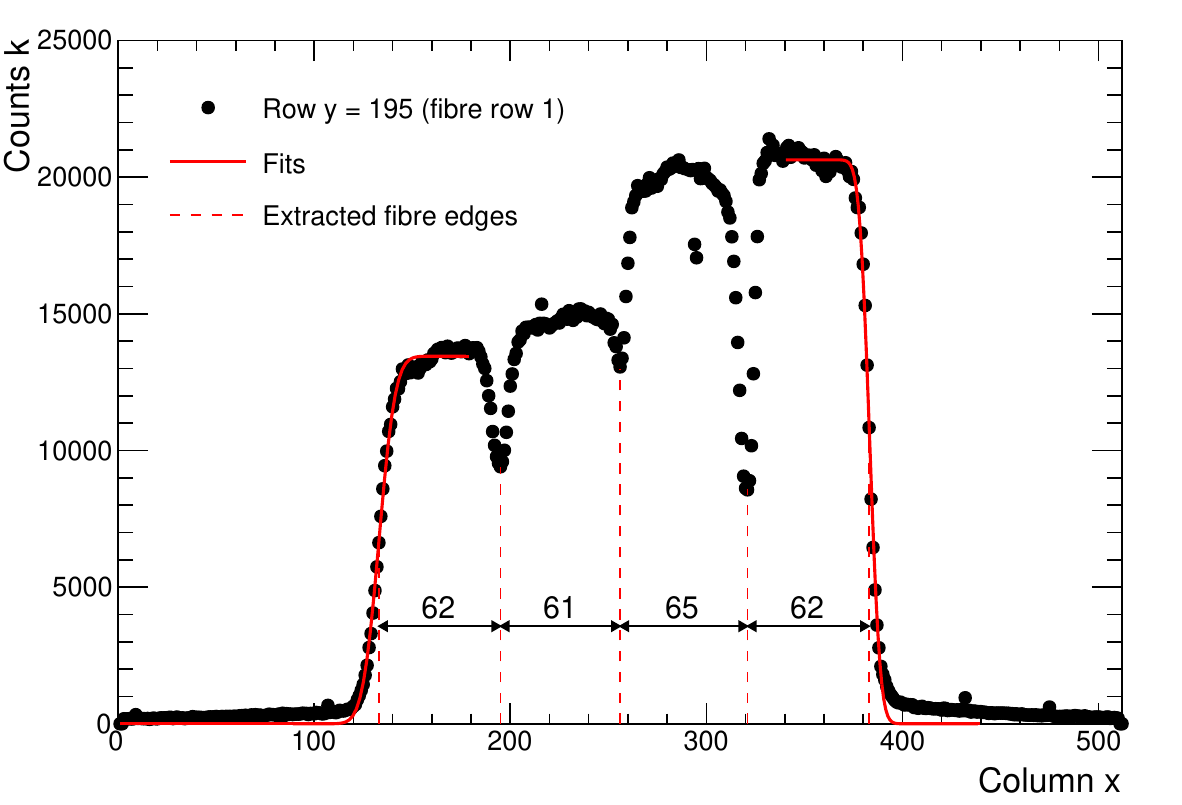}
        \caption{}
        \label{fig:UV_columns_cross-section}
    \end{subfigure}
    \caption{(a) Images created by summing 2,560 frames with the collimated blue light source positioned over each fibre column. (b) Slice of image of fibre bundle created by summing all four images in (a). Fits are used to extract the edges of the fibre bundle and the minima inside these edges are used as the fibre-to-fibre interfaces.}
\end{figure*}

The images in Fig.~\ref{fig:UV_columns_individual} show the 2D projection of the SciFi bundle on to SwissSPAD2. 
The contour of each fibre cross-sectional profile was obtained by fitting the signal count distribution with an error function, as shown in Fig.~\ref{fig:UV_columns_cross-section}.
During the tracking analysis, the contour positions will be used to identify which fibre was crossed by an electron and the centre of the fibre will identify the particle 2D position. 

\subsubsection*{Masking noisy pixels}
The unique feature of SPAD array sensors is the possibility to handle each single pixel independently and exploit both its time and position information. 
Such a feature allows to disable noisy or problematic pixels either on the hardware, firmware or simply during the image analysis, depending on the particular design of the chip. This feature was not implemented into \spadwithoutlens, therefore in the measurements reported later in Sec.~\ref{sec:results-without-mulens},~\ref{sec:data-analysis-imaging},~\ref{sec:test-scifi-sipm} and~\ref{sec:mat-saint-gobain} it was sufficient to mask the noisy pixels at the stage of the single frame analysis. 

Since the goal is to detect events where only a few signal counts are expected, it becomes crucial to minimise the contribution of dark counts (DC), whilst keeping a sufficiently-high PDE. 
This would ensure a good tracking efficiency as well as a low probability to randomly misidentify distributed DC as a track (fake track). It is worth noting that disabling pixels reduces the number of DCs as well as the effective fill factor. Hence, a compromise has to be found.

\begin{figure*}
    \centering
    \begin{subfigure}[b]{0.5\textwidth}
        \includegraphics[width=\linewidth]{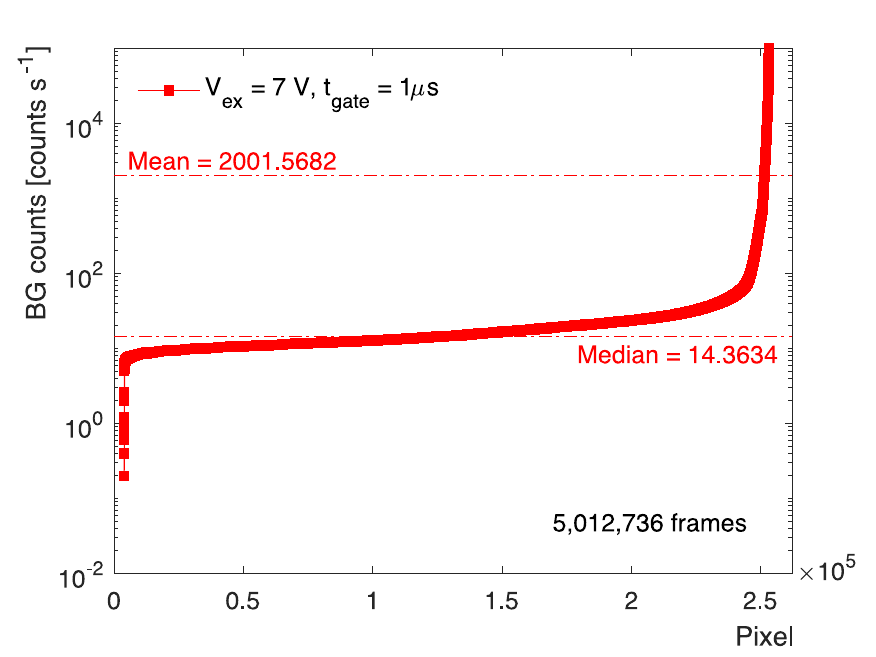}
        \caption{}
        \label{fig:BG_perPixel}   
    \end{subfigure}%
    \begin{subfigure}[b]{0.5\textwidth}
        \includegraphics[width=\linewidth]{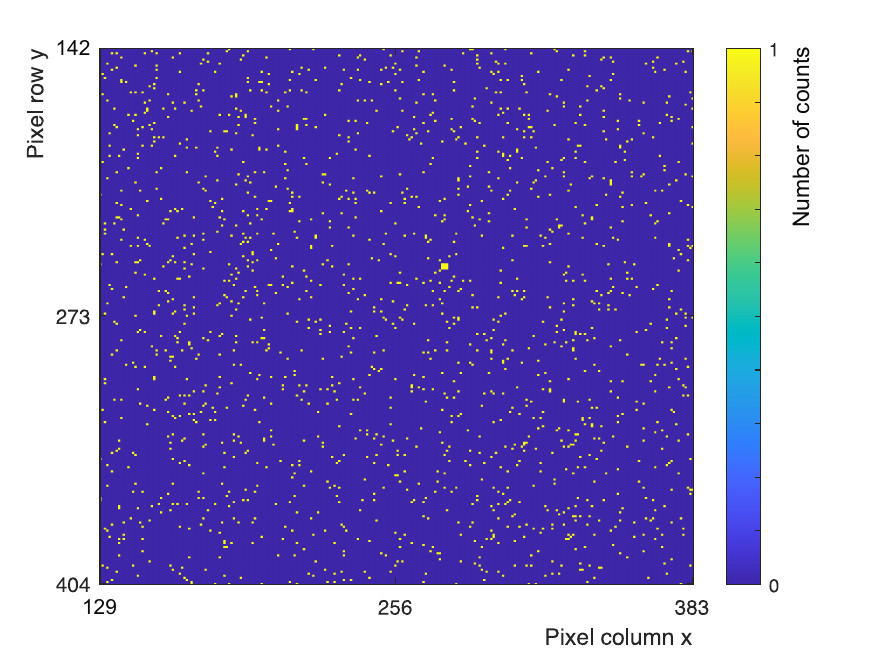}
        \caption{}
        \label{fig:BG_map}
    \end{subfigure}
    \caption{
    (a) Total counts seen in each pixel for $>5~\text{M}$ frames in cps. For the measurement, the noisiest top $5\%$ of pixels were masked.
    (b) Distribution of the masked pixels in the fibre bundle region.
    }
\end{figure*}

Over 5,000,000 frames were summed and the total number of counts per pixel was plotted as a function of the pixel number. The pixels were ordered from lowest to highest DC rate (DCR), as shown in Fig.~\ref{fig:BG_perPixel}.
The large variation in DCR per pixel is due to different defect densities pixel to pixel which arise due to non-uniformities in CMOS fabrication processes.
The median DCR was measured to be just over 14 counts per second (cps). The noisiest top $5\%$ of pixels, i.e. with a DCR above 117~cps, were not used in the data frame analysis of the subsequent measurements. The distribution of the masked pixels in the fibre bundle region is quite uniform as can be observed in Fig.~\ref{fig:BG_map}. A study measuring the cross-talk probability between individual SPAD pixels performed \cite{UlkuThesis} showed that the integral of the nearest neighbour pixel cross-talk is less than 0.5$\%$ sample to sample. 
One expects that 0.6 unmasked pixels are triggered through cross-talk by one of the masked pixels per frame.

\subsection{Impact of pixel microlenses on SwissSPAD2 fill factor}
\label{sec:results-without-mulens}
A first set of measurements with \SrY~electrons was performed with \spadwithoutlens. The single-frame configuration with a $1~\mu\text{s}$ gate allowed to detect and track single electrons produced by \SrY~with a relatively small contribution from dark counts. The 7~V excess bias provided the maximum PDE, i.e. slightly above 4\% at the scintillating fibre emission peak (450~nm). 2,981,888 frames were collected to characterise the DCR. Then, the same number of frames was collected while exposing the SciFi bundle to \SrY~electrons.

A second round of measurements was performed with \spadwithlens. This time 5,012,736 frames where captured in each of the two data runs, one for the DCR characterisation and one to detect \SrY~electrons.

\begin{figure}
    \centering
    \includegraphics[clip, trim=3cm 9cm 3cm 9cm, width=0.98\linewidth]{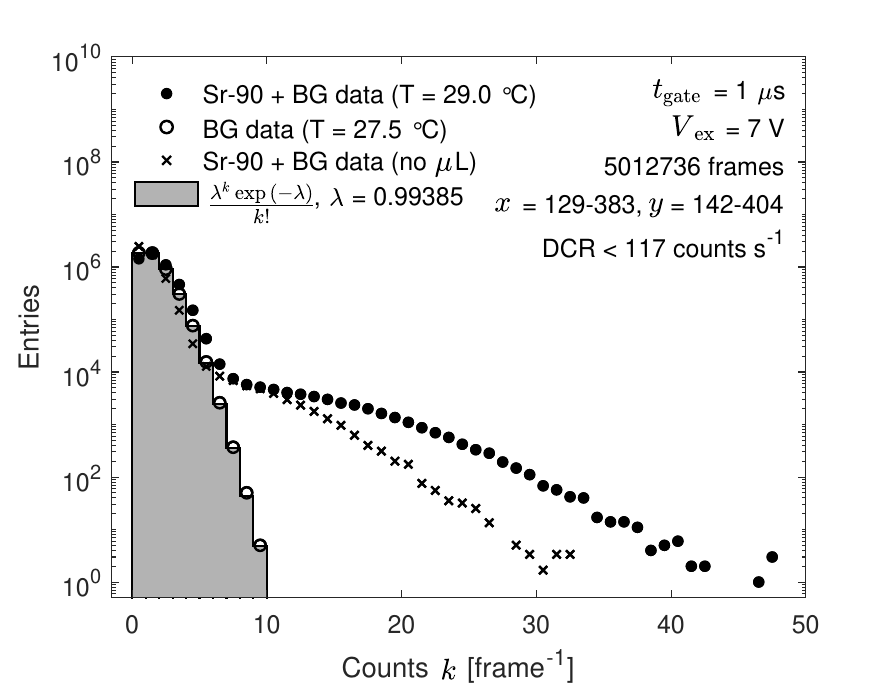}
    \caption{Distribution of number of counts per $1~\mu\text{s}$ frame corresponding to the fibre bundle region on SwissSPAD2. 
    The hollow markers show the distribution obtained from data collected without \SrY~source with \spadwithlens, known as the background (BG) data. 
    The grey histogram is a Poisson distribution with $\lambda$ equal to mean counts of the BG data.
    The solid dots show the distribution of the number of counts per frame taken with the \SrY~source positioned above the fibre bundle with  \spadwithlens~and the crosses show data obtained with \spadwithoutlens~without microlenses.
    The data collected with \spadwithoutlens~without microlenses
    and the Poisson distribution were normalised to 5,012,736 frames.
    } 
    \label{fig:BG_frame_counts_distribution}   
\end{figure}

Fig.~\ref{fig:BG_frame_counts_distribution} shows the distribution of the number of counts per $1~\mu\text{s}$ frame in the fibre bundle region after disabling the noisy pixels. The measurement taken in absence of \SrY~source provides the system background (BG) (hollow markers), dominated by DCs. The data follows closely a Poisson distribution, shown with a grey histogram. The distribution of the number of counts is also shown for data collected while exposing the fibre bundle to \SrY~electroncs for both \spadwithlens~(solid dots) and~\spadwithoutlens~(crosses). In both cases, the number of counts reaches a value definitely higher than for BG data, a clear signature of electrons traversing the fibre bundle. The slight shift in the peak of the number of counts per frame with respect to the BG data is simply given by the simultaneous contribution of BG and signal electron counts. Moreover, slightly different environmental conditions can also introduce additional small differences in the DCR. For instance, a difference in ambient temperature between the two measurements was recorded ($27.5^{\circ}\text{C}$ for the BG data and $29.0^{\circ}\text{C}$ for the \SrY~data).

As anticipated above, although the fill factor could theoretically increase up to a factor five for photons perpendicular to the SPAD surface, it is worth noting that scintillation photons exit the fibre end with a certain angular distribution~\cite{kuraray-catalogue-ps}. Moreover, systematic effects like microlens-SPAD misalignment can further reduce the effective fill factor. Hence, the impact of pixel microlenses on the SPAD PDE shall be studied
directly on data. Since \spadwithlens~and~\spadwithoutlens~were exposed to the same type of electrons (energy, direction, etc.), the comparison between the position of the respective spectrum tails
allows to claim a relative increase in PDE of around 50\%, up to an absolute PDE of approximately 6\% for \spadwithlens~around the fibre scintillation emission peak.

\subsection{Tracking of \SrY~electrons}
\label{sec:data-analysis-imaging}
The demonstration of particle tracking with electrons from \SrY~was performed with \spadwithlens, given the higher PDE. The goal is to detect unambiguous tracks of electrons in a bundle of scintillating fibres read out with a SPAD sensor. To our knowledge, such a measurement has never been reported in literature yet.

A simple algorithm was developed to automatically scan through all frames and search for vertical track-like events. 
The pixel-to-fibre mapping, extracted in Sec.~\ref{sec:test-characterisation}, was used to count the number of photons detected from each fibre independently. For this study, a track is defined by three parameters. Firstly, the number of fibres that the electron travels through, secondly the number of counts per track, and thirdly the number of counts per fibre. Since the \SrY~source is positioned at the top of the fibre bundle, an electron is assumed to always pass through the first row of fibres. The algorithm loops through all frames and counts the number of hits detected in each fibre region in the first row. If there is at least one count in any of the fibres, the algorithm checks the fibres directly below in the same column. If each row also contains at least one count, the frame is considered as a track-like event. This is repeated for each column. To remove frames that may contain multiple tracks, electrons that scatter and travel through multiple columns, or simply frames that contain too much noise, an SNR analysis was performed. 
All counts in the track fibre regions are considered signal and everything else is considered noise. If the $\text{SNR}>1$, the frame is stored as a track-like event. 

The same track selection algorithm was performed on both the BG and $^{90}$Sr+BG datasets to assess the type and number of fake tracks that arise due to DCs. This was performed using multiple track selection parameters and is summarised in Tab.~\ref{tab:Tracks_1us}. 
It was found that for events with three fibre hits the probability that in \SrY~data an electron fake-track was generated from DC is less than 2\% with at least four counts. Similarly, if four fibres were hit, the fake-track probability was less than 3\% with at least four counts. The fake electron track probability becomes negligible if just a few more counts are required. It is interesting to note  that, simply by considering combinatorial probability, the chance to misidentify a particle track from dark counts further reduces near proportionally with the smaller diameter of the scintillating fibre. 

\begin{table*}
    \centering
    \caption{Summary of track selection performed on BG and \SrY+BG datasets. The number of tracks in the table are normalised to same number of frames (5,012,736) for both the BG and \SrY+BG data.}
    \begin{tabular}{c|cc|cc|c}
        \hline
        \multirow{2}*{No. fibres} & Min. counts & Min. counts &\multicolumn{2}{c|}{No. of tracks} & Misidentification \\ 
     & per track & per fibre & BG & $^{90}$Sr+BG & probability \\
        \hline
        \multirow{5}*{3}    & $3$ & $1$ & $3911$ & $28108$ & $13.9$ \\
                            & $4$ & $1$ & $363$ & $20808$ & $1.7$ \\
                            & $5$ & $1$ & $23$ & $19055$ & $0.1$ \\
                            & $6$ & $1$ & $0$ & $17684$ & $0$ \\
                            & $6$ & $2$ & $0$ & $8160$ & $0$ \\
        \hline
        \multirow{5}*{4}    & $4$ & $1$ & $231$ & $8372$ & $2.8$ \\
                            & $5$ & $1$ & $31$ & $7848$ & $0.4$ \\
                            & $6$ & $1$ & $1$ & $7607$ & $>0.1$ \\
                            & $6$ & $2$ & $0$ & $2338$ & $0$ \\
        \hline
    \end{tabular}
    \label{tab:Tracks_1us}
\end{table*}

\begin{figure*}
    \centering
       \begin{subfigure}{0.32\textwidth}    
            \includegraphics[width=\linewidth]{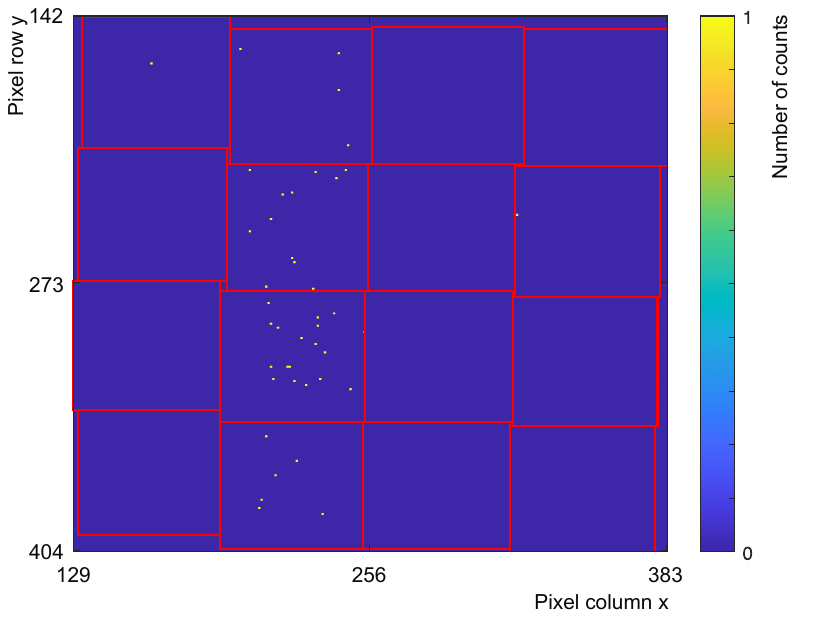} 
            \includegraphics[width=\linewidth]{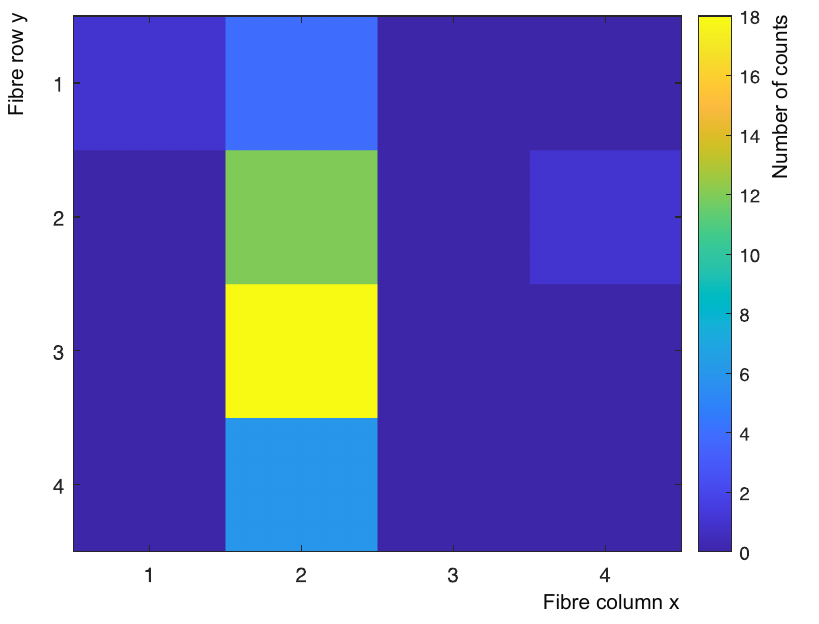}
            \caption{}
            \label{fig:Tracks_Sr90_throughgoing}
        \end{subfigure}
        \begin{subfigure}{0.32\textwidth}
            \includegraphics[width=\linewidth]{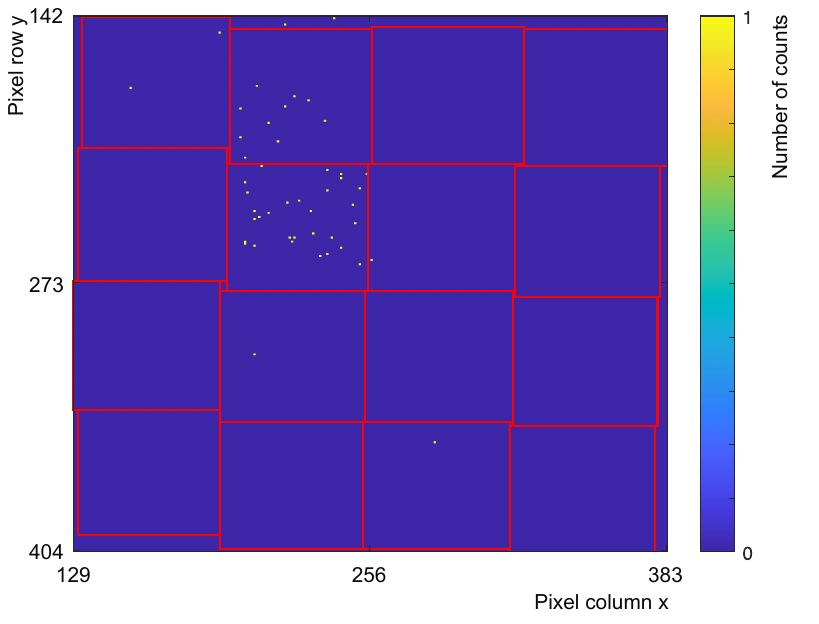}
            \includegraphics[width=\linewidth]{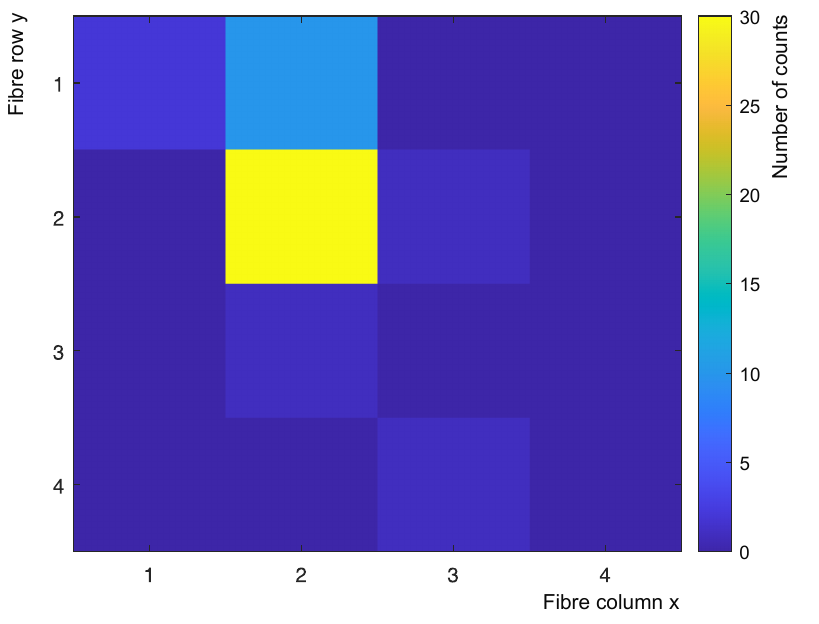}
            \caption{}
            \label{fig:Tracks_Sr90_stopping}
        \end{subfigure}
        \begin{subfigure}{0.32\textwidth}
            \includegraphics[width=\linewidth]{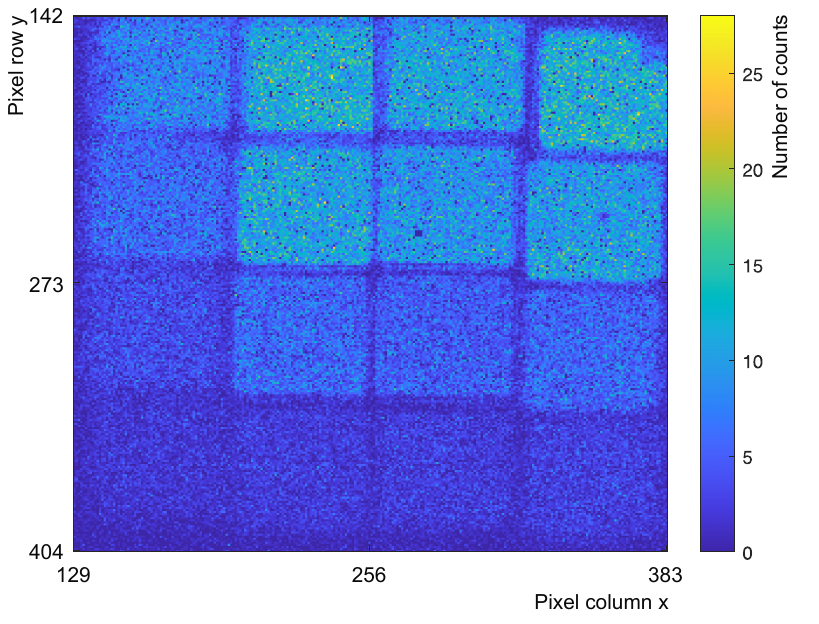}
            \includegraphics[width=\linewidth]{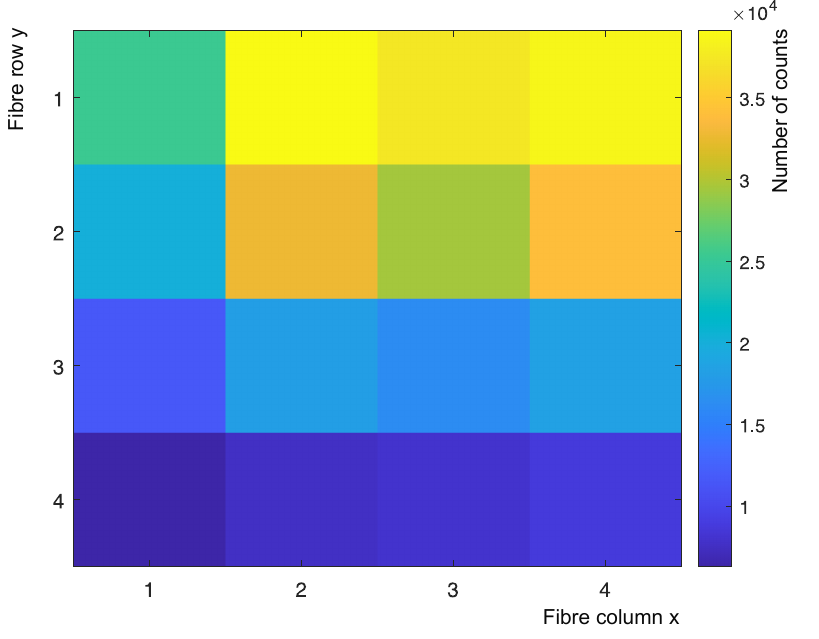}
            \caption{}
            \label{fig:Tracks_Sr90_flux}
        \end{subfigure}
    \caption{
    Two candidate electron tracks from the \SrY~dataset collected with \spadwithlens~are shown, respectively a vertical through-going track in \ref{fig:Tracks_Sr90_throughgoing} and a vertical stopping track in \ref{fig:Tracks_Sr90_stopping}. The top figures show the counts in the $4 \times 4$ fibre region on \spadwithlens~with single-pixel binning. The red lines show the fibre contours computed as in Sec.~\ref{sec:test-characterisation}. The bottom figures show the same events regrouped with single-fibre binning. In Fig. \ref{fig:Tracks_Sr90_flux} the flux of \SrY~electrons selected from the vertical-track analysis is shown. Some non-uniformities between fibres is likely due to slightly different quality in the treatment of the fibre end.
    }
    \label{fig:Tracks_Sr90}
\end{figure*}

Examples of track-like events from the \SrY~source dataset can be seen in Fig.~\ref{fig:Tracks_Sr90}. In the same figure, also the sum of the counts observed from all the selected vertical tracks is shown.

\subsection{Comparison with SiPM readout}
\label{sec:test-scifi-sipm}
A subset of the data acquired with \spadwithlens, where electrons travel vertically through two fibres and come to rest in the third, have been compared to the same type of events captured by SiPMs.

One Hamamatsu S13360-1350CS MPPC with 50~$\mu$m pitch and maximum nominal PDE = 40\% at 450~nm \cite{hamamatsu:mppc},
was precisely coupled to each of the sixteen fibre ends thanks to a 3D printed black optical connector. Each connector contains a $1~\text{mm}$ thick foam acting like a spring, pushing the SiPM against the fibre end to minimise the relative distance. 

The sixteen SiPMs were connected to individual channels of a 32-channel CAEN DT5702 readout board \cite{caen-catalogue}. The same \SrY~electron source was positioned directly above the fibre bundle as before. However, rather than obtaining the list of pixels with a detected photon in binary format as with the SwissSPAD2 sensors, each SiPM generates an analogue signal whose height is proportional to the number of photons detected per event in the whole SiPM active area. Data was acquired for $150\,\text{s}.$ 

The SiPM signal is given in units of Analogue-to-Digital Converter (ADC). The pedestal, between 100 and 150 ADC, was subtracted and the gain was extracted as the conversion factor from the measured number of signal ADCs to the corresponding number of photoelectrons (PE) by measuring the distance between the single-PE peaks visible below 500 ADCs. An average gain of 44 ADC/PE was obtained.

In the SiPM setup a trigger was implemented at the electronics level by requiring signal in at least one of the fibres on the third layer from top. Such a trigger was useful to reduce the pile-up of \SrY~electrons and avoid any event loss during the data taking.

\begin{figure}
    \centering
    \includegraphics[width=\linewidth]{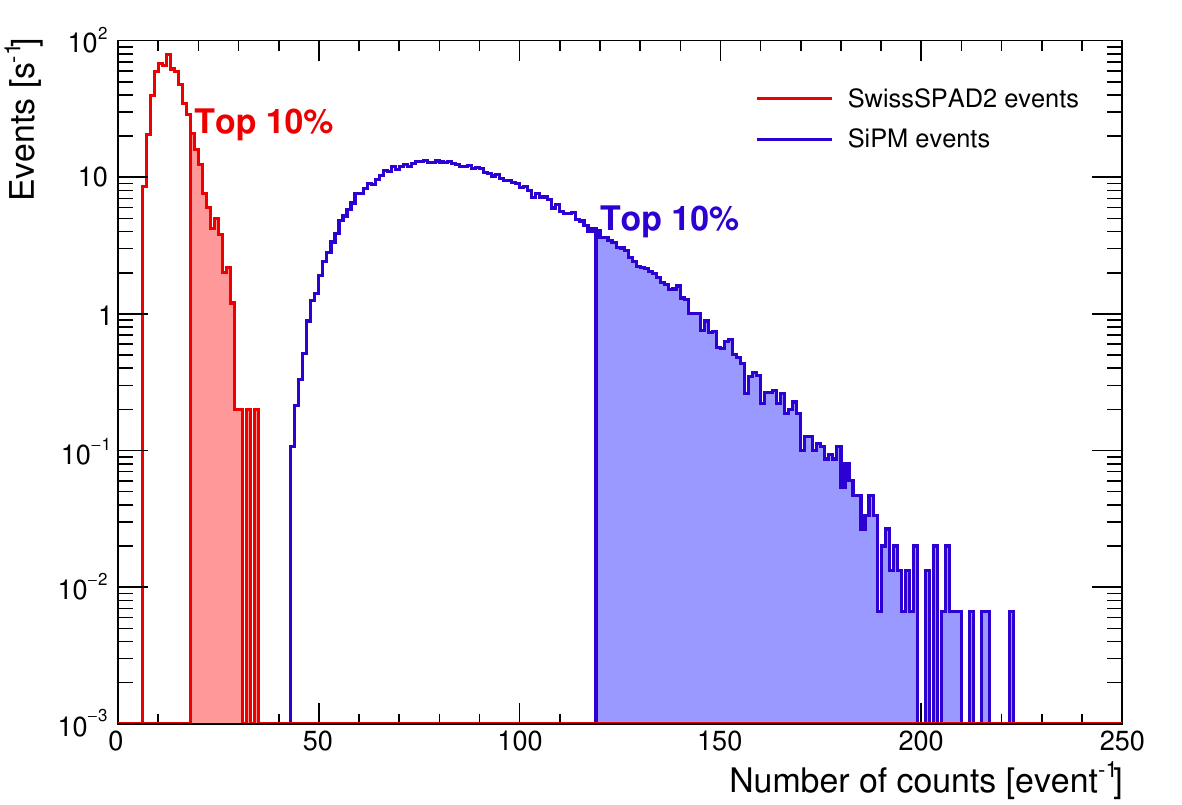}
    \caption{
    \label{fig:SiPM_SPAD_comparison}
    Distribution of the number of counts measured in \spadwithlens~(red) and SiPMs (blue) obtained by exposing the fibre bundle to \SrY~electrons. Each event corresponds to a selected vertical track where electrons have travelled through two fibres and come to rest in the third. The two distributions are normalised to the number of events for one second exposure and are obtained after a cut on the minimum number of counts per fibre has been applied to provide about the same number of selected events (655 for \spadwithlens~and 687 for SiPMs).
    } 
\end{figure}

Vertical, stopping tracks were selected from the tracks identified in Sec.~\ref{sec:data-analysis-imaging}. Exactly the same selection has been applied to the events detected with SiPMs in order to have a direct comparison with \spadwithlens~response to \SrY~electrons.
However, given the different threshold of \spadwithlens~and SiPMs, different cuts on the minimum number of counts per fibre were applied to obtain approximately the same number of selected events. 
This approach avoids to bias the selection that, otherwise, would be affected by the different PDE values. Requiring at least 2 (14) counts in \spadwithlens~(SiPM) per fibre provided 655 (687) selected events for one second of exposure. The distributions of the total number of counts per event per second can be seen in Fig.~\ref{fig:SiPM_SPAD_comparison}.

Given that the tail ends of both distributions correspond to similar events, a comparison of the number of detected photons provides an approximate estimate of \spadwithlens~PDE relative the SiPM one. In order to limit the impact of statistical fluctuations, the relative \spadwithlens~and SiPM PDEs were obtained by integrating the spectrum tail until the right-most 10\% of the events were accounted for. The corresponding ratio of the obtained number of counts per fibre was used to estimate the ratio between the \spadwithlens~and SiPM overall PDE values, that was found to be about 6 times lower. The same result was obtained also by changing the spectrum tail integration range. This value is approximately in agreement with the SwissSPAD2 PDE \cite{SwissSPAD2Ulku2019} as well as with the increase in absolute PDE given by the pixel microlenses, quantified to be about $6\%$ at 450~nm in Sec.~\ref{sec:results-without-mulens}.

The measured signal extrapolated to a minimum ionising particle (2 MeV/cm) crossing a 1~mm thick scintillating fiber (see Sec.~\ref{sec:scifi-simulation} for details on the simulation) would correspond to approximately 25 signal counts for the used SiPM and about 3.5 counts for \spadwithlens.

\subsection{Setup with scintillating fibre ribbons}
\label{sec:mat-saint-gobain}
As mentioned above, an optimal solution for a SciFi detector capable of 3D particle tracking by detecting photons produced in the scintillating fibres and isotropically emitted, whilst keeping the single-fibre resolution in every spatial coordinate, is to alternate one-fibre-thick layers with different orientation (e.g. +45/-45/+45/-45/... or 0/+60/-60/0/+60/-60...) resulting in an XY structure.

\begin{figure*}
    \centering
       \begin{subfigure}{0.32\textwidth}    
            \includegraphics[width=\linewidth, clip, trim=1cm 1cm 1cm 1cm]{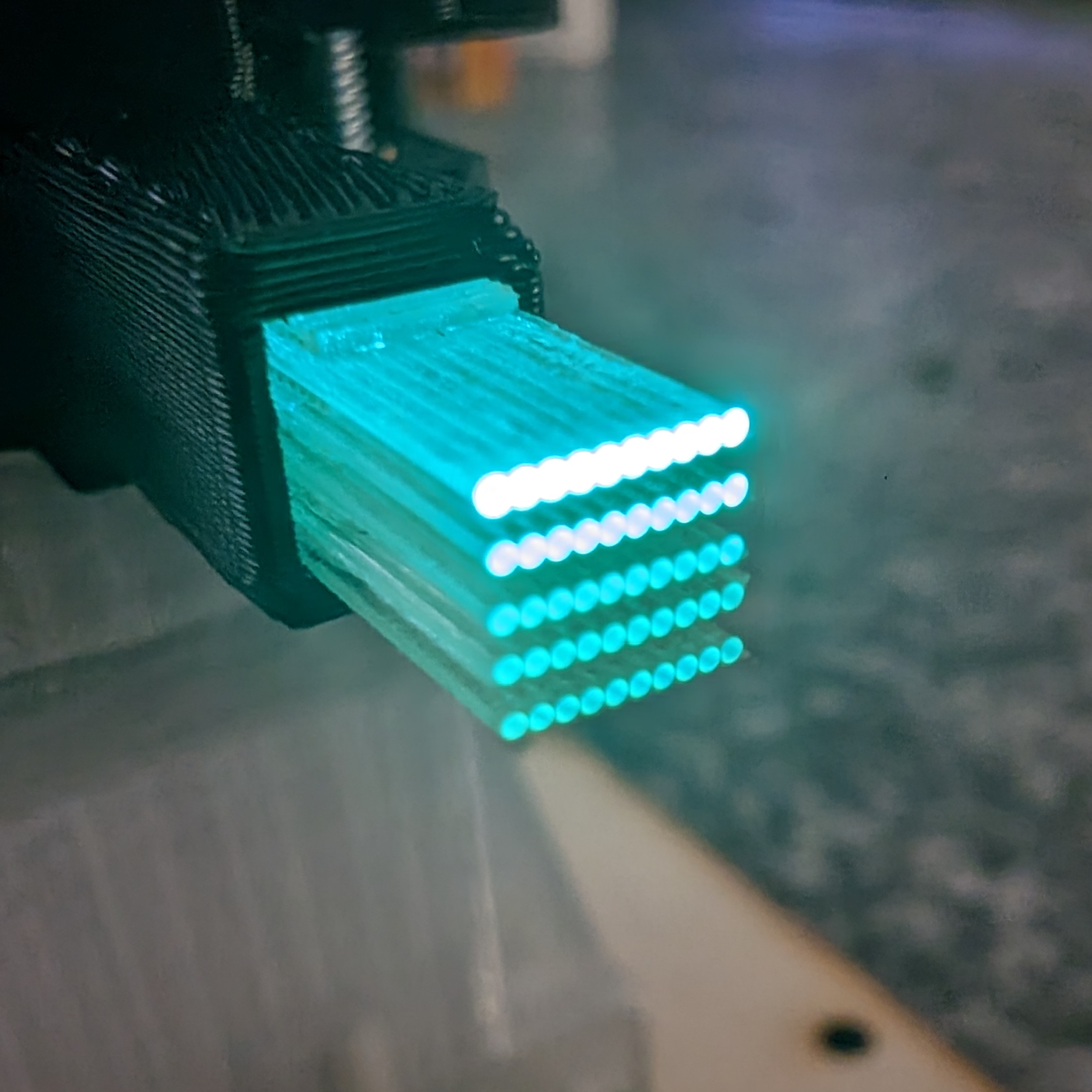} 
            \caption{}
            \label{fig:XYBundle_UV}
        \end{subfigure}
        \begin{subfigure}{0.32\textwidth}
            \includegraphics[width=\linewidth, clip, trim=18cm 18cm 18cm 18cm]{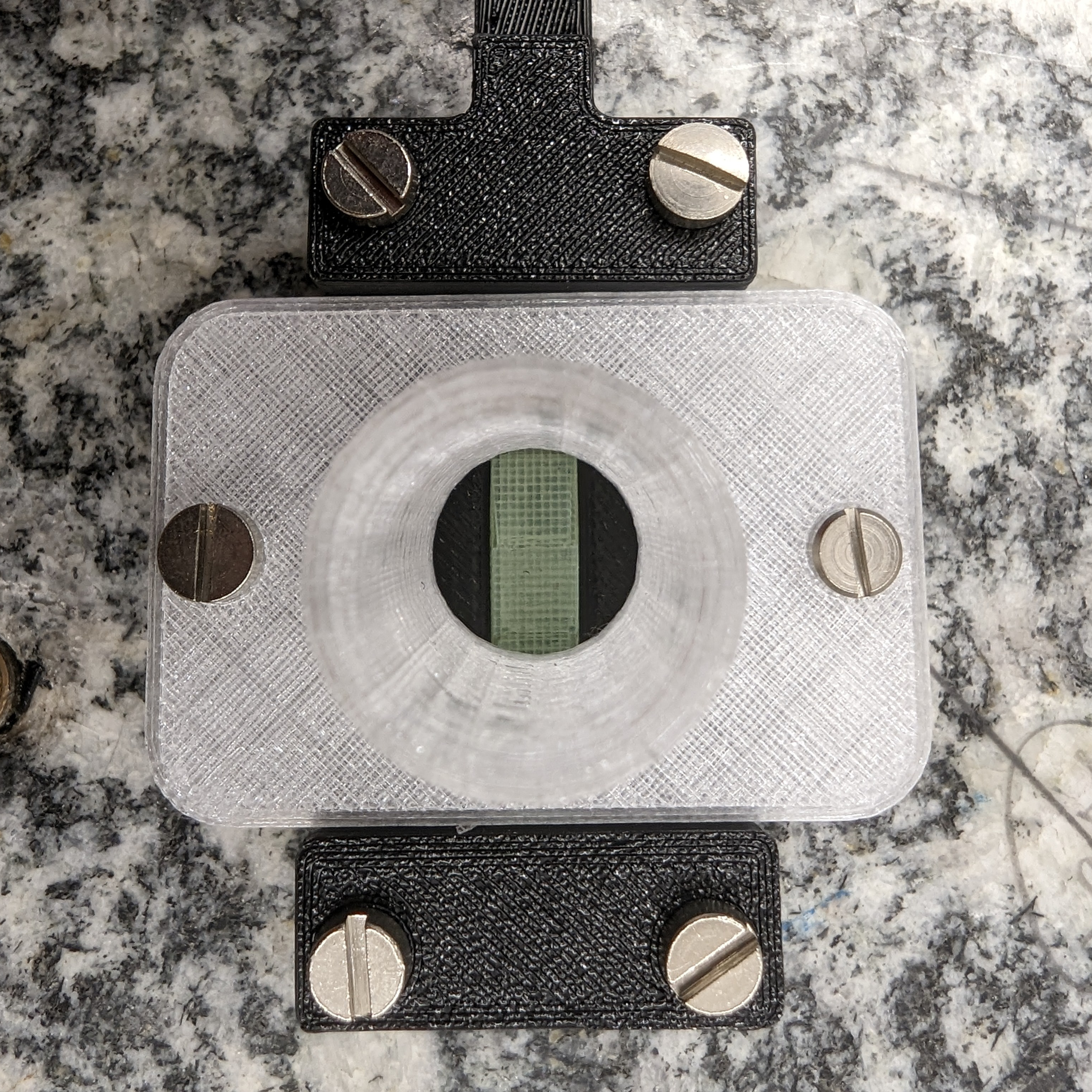}
            \caption{}
            \label{fig:XYBundle_TopView}
        \end{subfigure}
        \begin{subfigure}{0.32\textwidth}
            \includegraphics[width=\linewidth, clip, trim=1cm 1cm 1cm 1cm]{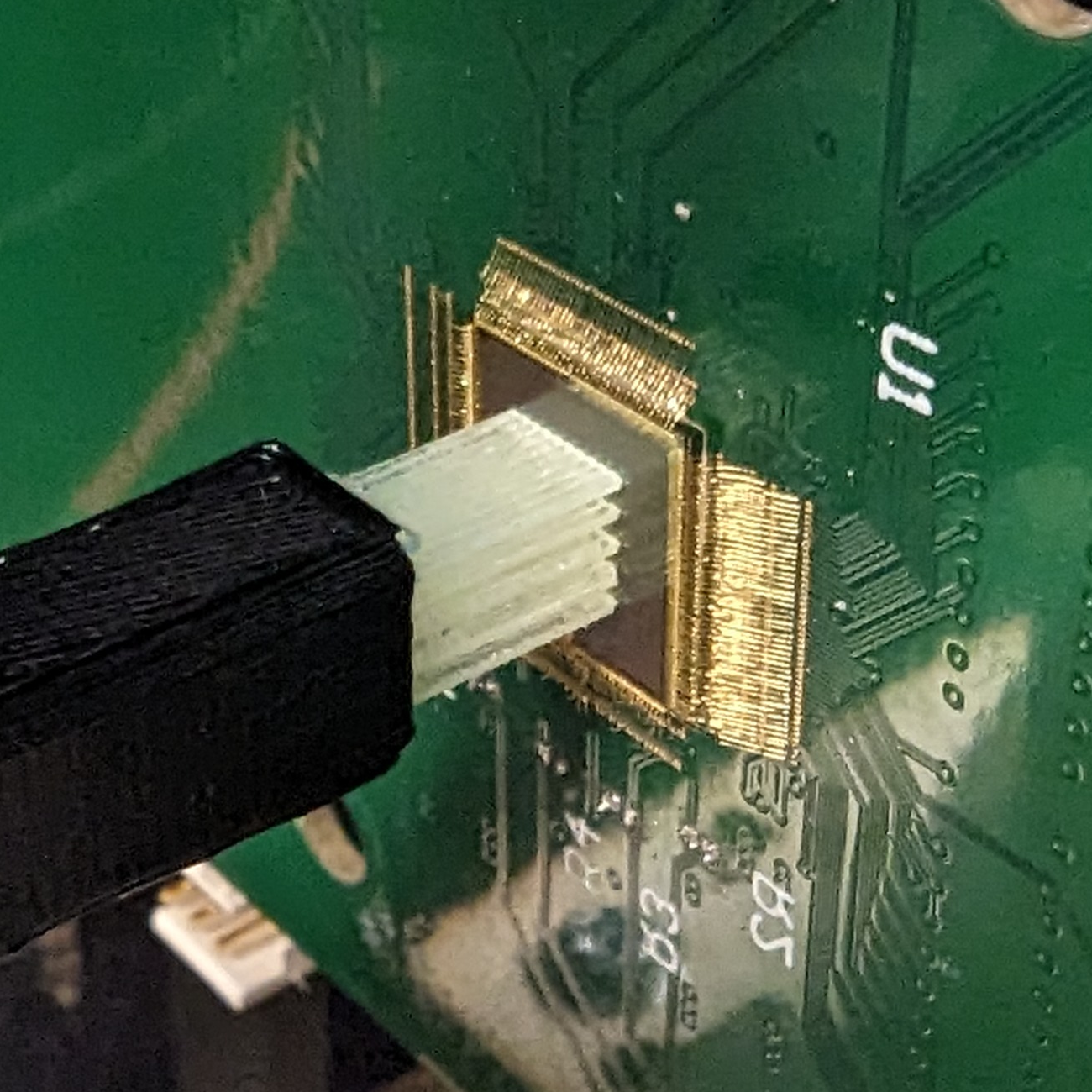}
            \caption{}
            \label{fig:XYBundle_Couple}
        \end{subfigure}
    \caption{
    (a) Close-up of XY fibre bundle module illuminated with blue light source. 
    (b) Top-down view of XY fibre bundle to illustrate perpendicular fibre layers.
    (c) XY fibre bundle module pointed at \spadwithlens.
    }
\end{figure*}

A small module compatible with the above setup was constructed to perform initial tests on a demonstrator with an XY structure. The prototype is shown in Fig.~\ref{fig:XYBundle_UV}, illuminated from the top with the blue light source. The dimensions of the module are identical to those of the clamp described above, except that it features a $5\times5~\text{mm}^{2}$ channel to accommodate different scintillating fibres.

For this demonstrator, BCF-20 fibre \textit{ribbons} produced by Luxium Solutions (formerly Saint-Gobain Crystals)~\cite{saint-gobain} were used. The ribbons were produced as ten, $1~\text{m}$ long, $500~\mu\text{m}$ diameter fibres with a circular cross-section, glued together side-by-side. Also, unlike the fibres discussed in the previous sections, these fibres emit green light with an emission wavelength of $492~\text{nm}$, which in principle better suits the peak PDP of the SwissSPAD2 sensor.

Similarly to the previous setup, only one plane is read out. For the fibre layers in the $x$ plane, a single ribbon was cut into five $8~\text{cm}$ long pieces. For fibre layers in the $y$ plane, smaller $5~\text{mm}$ long pieces of ribbon were cut and glued using EJ-500 optical cement side-by-side, perpendicular to the longer lengths. This structure simply fakes the spacing between layers in an XY configuration. Moreover, it helped to test on a small scale the gluing of multiple layers before moving to a bigger prototype. This solid volume of fibre material below the position of the \SrY~source can be seen in Fig. \ref{fig:XYBundle_TopView} from the point of view of the source.

\begin{figure*}
    \centering
        \begin{subfigure}{0.49\textwidth}
            \includegraphics[width=\linewidth]{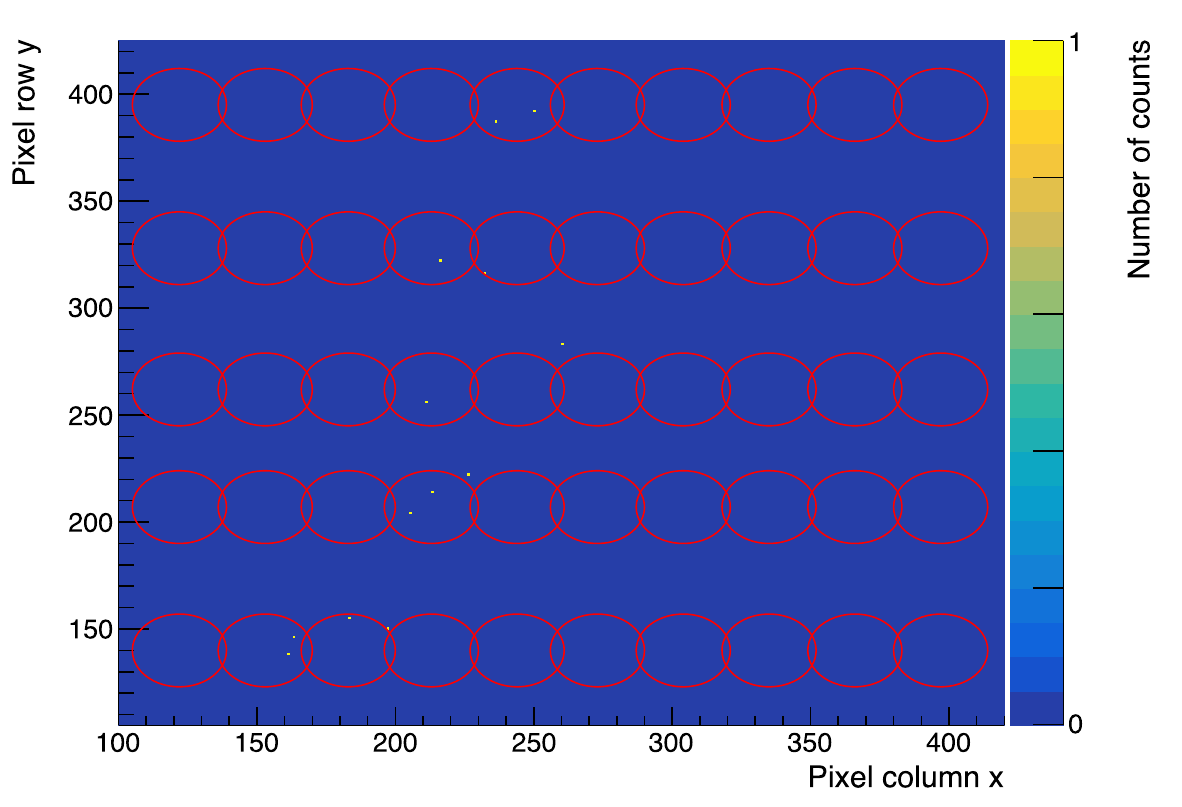}
            \caption{}
            \label{fig:XYBundle_TH2}
        \end{subfigure}
        \begin{subfigure}{0.49\textwidth}    
            \includegraphics[width=\linewidth]{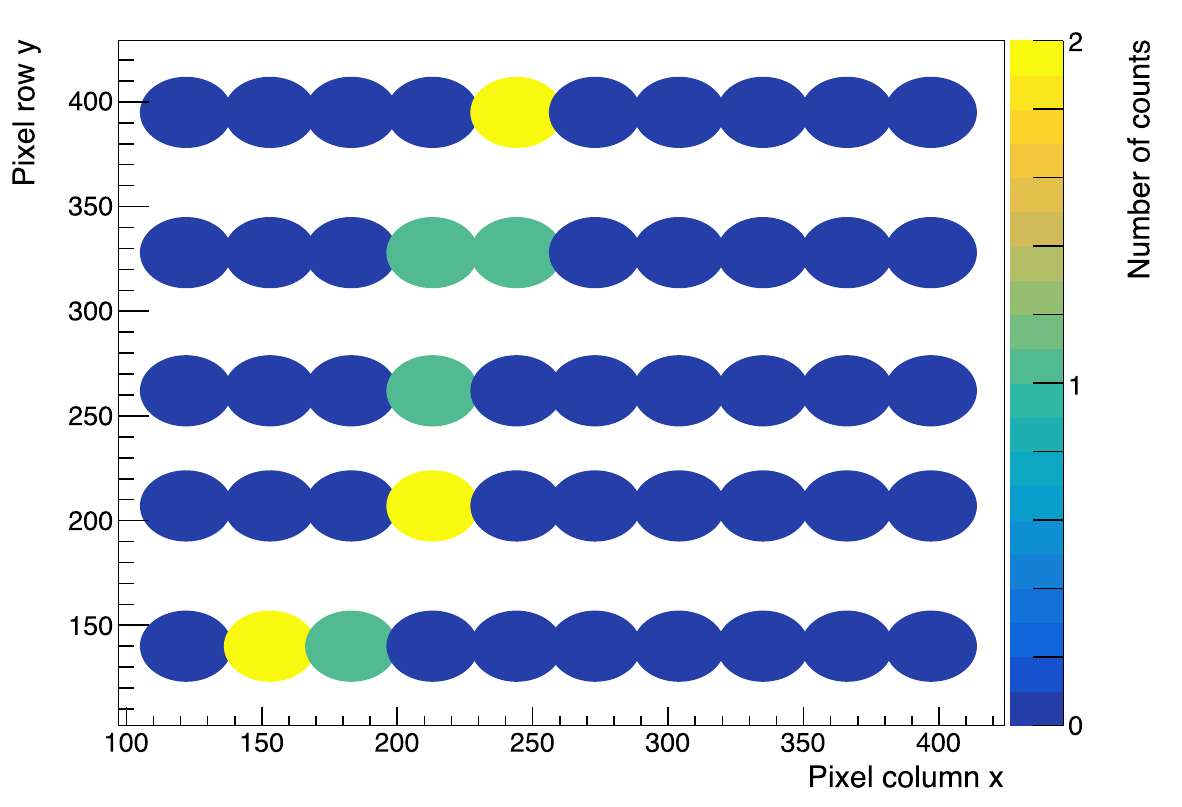}
            \caption{}
            \label{fig:XYBundle_TH2Poly}
        \end{subfigure}
    \caption{
    (a) Candidate electron track with single-pixel binning.
    (b) Candidate electron track with single-fibre binning. The colour of the circles represents the number of counts per fibre. The three counts falling outside the fibre contours (see Fig.~(a)) are not included in the image.
    }
\end{figure*}

As before, the fibre ends were polished. The \spadwithlens~was placed at a distance of less than $1~\text{mm}$ from the fibre ends as displayed in Fig. \ref{fig:XYBundle_Couple}. The same analysis method described above was performed again for this module. Fig. \ref{fig:XYBundle_TH2} shows a candidate electron track in a single $1~\mu\text{s}$ exposure frame with single-pixel binning, after masking noisy pixels and overlaying red circles showing computed fibre contours. The same frame is shown again in Fig. \ref{fig:XYBundle_TH2Poly} with counts regrouped into single-fibre bins.

Future R\&D plans include a second prototype with XY readout. 

\section{Scintillating fibres and SPAD arrays for precise neutrino detection}
\label{sec:scifi-simulation}
In the previous sections, the tests with SwissSPAD2 provided an initial demonstration of a novel particle detector configuration with scintillating fibres read out by SPAD array sensors.
These successful tests show the path towards a
drastic improvement in the technology of ultra-fine granularity scintillating detectors, thanks to the $\sim 10~\mu\text{m}$ spatial resolution of monolithic SPAD array sensors, combined with minimal deployment of electronics readout channels, which scales with the covered surface and no longer by the number of scintillating fibres.
It was also observed that the possibility to isolate the scintillation photons from a single particle event within a $\mu$s-scale time window combined with a fine granularity also minimises the probability to misidentify a particle track from dark counts, and that it would further reduce with the diameter of the scintillating fibres.
In addition, the sensors can be cooled, lowering the DCR by up to an order of magnitude for 20 degrees Kelvin~\cite{SPAD90nmHenderson,  SPAD150nmStoppa, SinglePhotonTimeRes10psCharbon}. To achieve miniature pixels, SwissSPAD2 was designed without timestamping on-chip, however, SPAD arrays with $\sim$ 50~ps single-photon timestamp resolution exist~\cite{SinglePhotonTimeRes10psCharbon}. This would allow the use of temporal correlation of hits to further reduce the amount of DCs inside the data sample without the need of using very short time gates.

A back-of-the-envelope calculation, based on the DCR of SwissSPAD2~\cite{SwissSPAD2Ulku2019}, results in an expected DCR of less than 2 DCs per 1000 250~$\mu$m fibres in a 10~ns time window at room temperature. Therefore, the background due to electronic noise can be considered negligible, which matches the measurements reported above.

In this section the application of this novel detector concept as well as its optimal configuration for the detection of neutrinos is studied using simulations. Such a setup would allow the detection of neutrino interactions with unprecedented spatial and time resolution.
The main advantages are expected for the following measurements, that can highly profit from a superb spatial resolution combined with an excellent particle identification: 
detection of low-momentum particles (below 300~MeV/c), such as protons or nuclear clusters, crucial for the understanding of nuclear effects;
improved charge and momentum reconstruction of final state particles;
high purity $\nu_e$ and $\bar{\nu}_e$ detection with minimal contamination from $\gamma \rightarrow e^+ e^-$ events, typically the dominant background;
and detection of final state neutrons, in particular low-energy ones that can deposit energy near the neutrino interaction vertex.
All these measurements will be crucial for the systematic reduction in the future LBL experiments.
Another application of such a novel detector configuration would be the detection of $\nu_{\tau}$ with precise identification of the $\tau$ decay vertex around 1 mm from its production point with timing information \cite{DONUT:2001}, which is, however, not studied in this work.

A setup assuming either an 6\% or 20\% overall PDE SPAD array sensor instrumenting a large active neutrino target was simulated. To achieve such a higher PDE one may: adopt fibres better matching to the SPAD peak PDP, as in Sec.~\ref{sec:mat-saint-gobain}; read out from both ends of the scintillating fibers; and improve the SPAD architecture by increasing the fill factor. The target consists of 2,000 XY-layers of 2 m long and 250 $\mu\text{m}$ thick fibres arranged in layers of 8,000 fibres each, resulting in a $2 \text{m} \times 2 \text{m} \times 1 \text{m}$ block. Using NEUT~\cite{NEUTSim2021}, 100,000 charged current muon neutrino ($\nu_\mu$) interactions following the T2K experiment energy flux~\cite{t2kfluxurl} were simulated. 
The final state particles were propagated through the detector geometry implemented in Geant4~\cite{Geant4RecentDevelopmAlliso2016, Geant4ASimulaAgosti2003, Geant4DevelopmAlliso2006}. The particle energy deposition was simulated and corrected for Birks' quenching~\cite{BIRKS196439}. The conversion from the deposited energy in the fibres (MeV) to the number of detected photons was validated by simulating and comparing with data from the \SrY~tests described in Sec.~\ref{sec:test-scifi-sipm}. Unfolding the 40\% PDE of the SiPMs and applying the 20\% (6\%) PDE of the assumed SPAD array sensors, an effective light yield of about 100 (30)~$\gamma$/MeV was determined.

\begin{figure*}
    \centering
       \begin{subfigure}{0.49\textwidth}    
            \includegraphics[width=\linewidth]{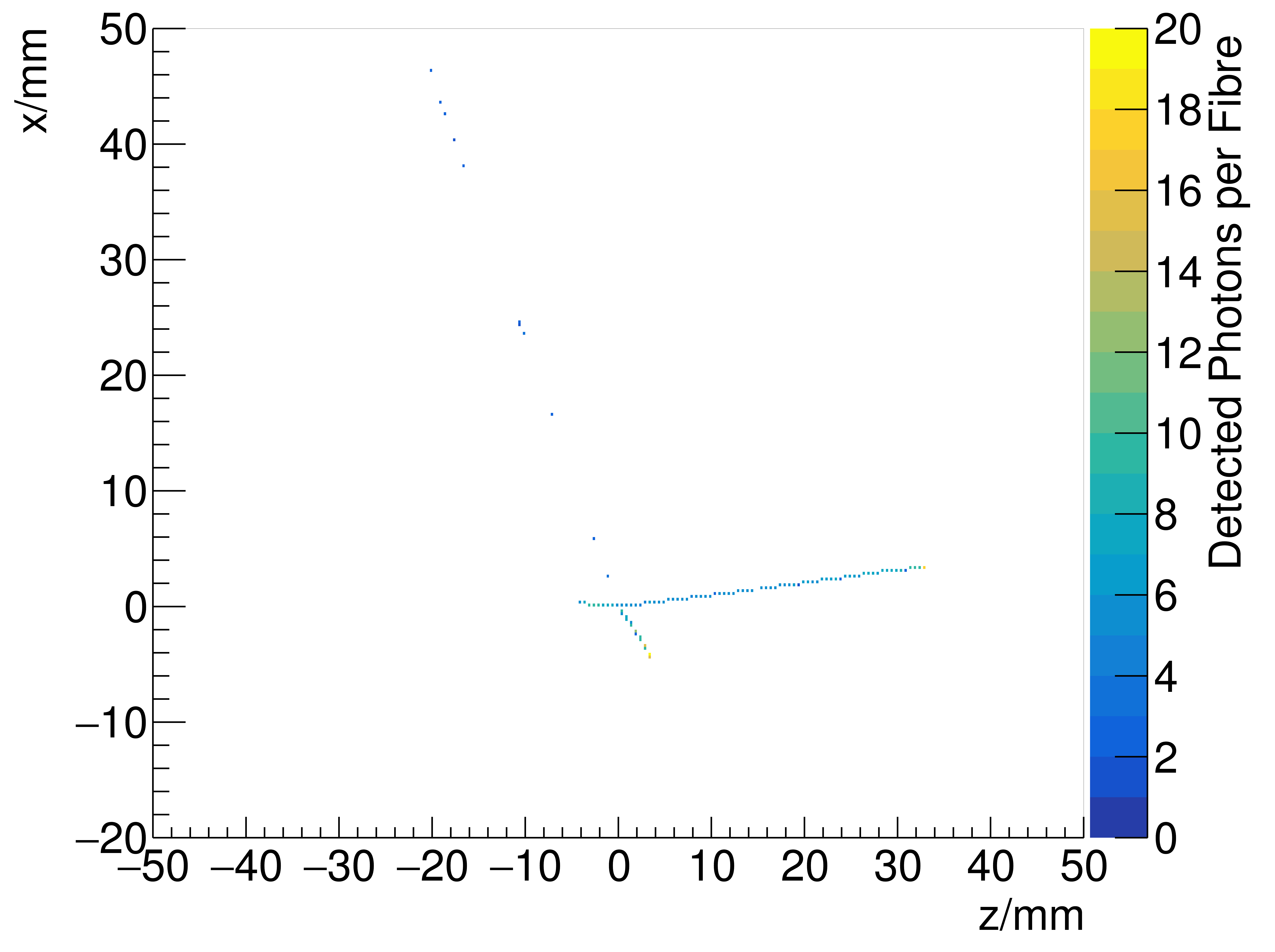} 
            \caption{}
            \label{fig:Sim_numu_CCQE_6pc}
        \end{subfigure}
        \begin{subfigure}{0.49\textwidth}    
                    \includegraphics[width=\linewidth]{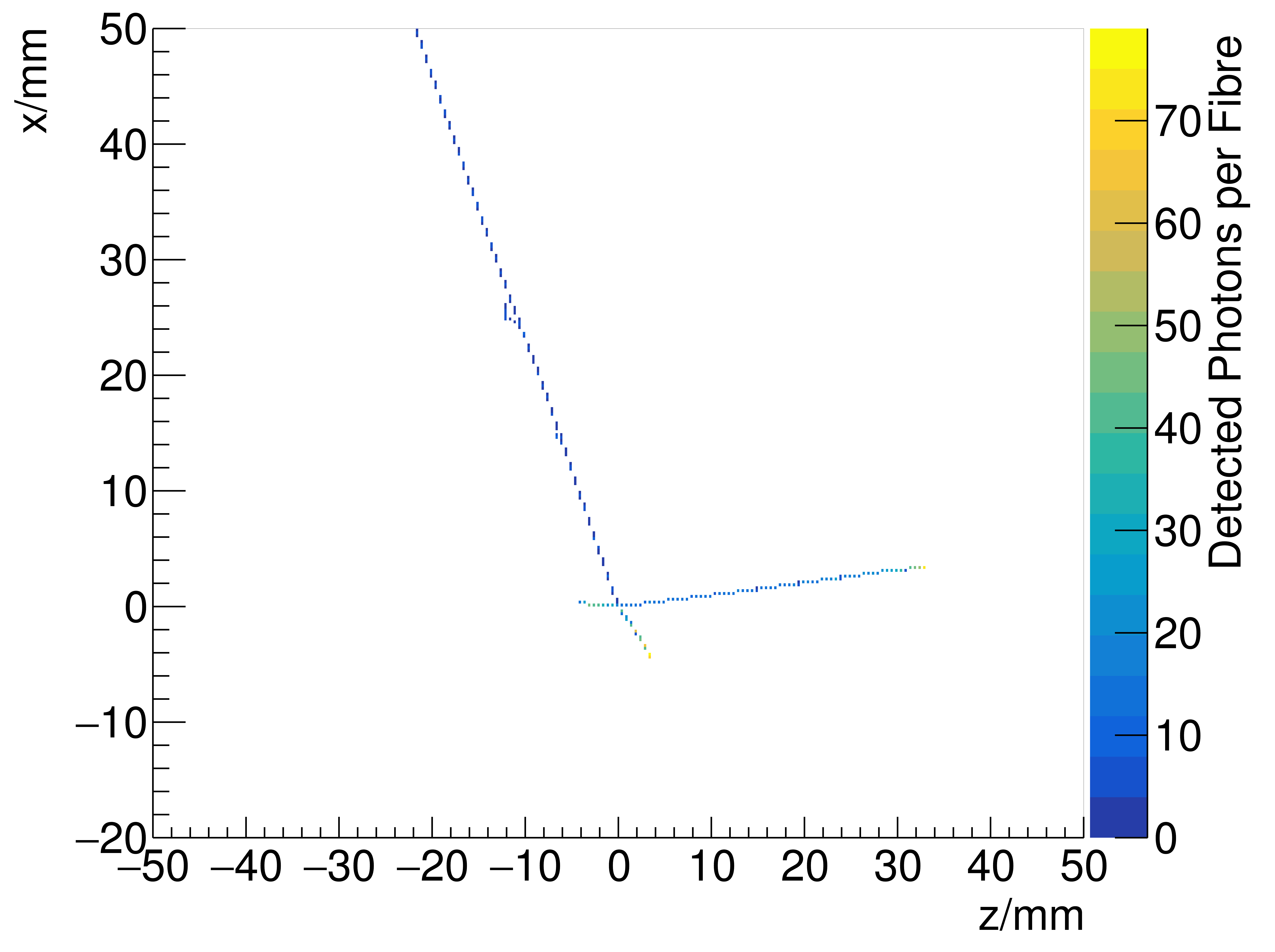} 
            \caption{}
            \label{fig:Sim_numu_CCQE_20pc}
        \end{subfigure}
    \caption{
    Simulated $\nu_{\mu}$ Charged Current Quasi Elastic (CCQE) interaction in the centre of the \scifi~detector, undergoing final state interactions, producing extra protons. The colour shows the number of photons detected from the scintillating fibres on the XZ projection with SPAD array sensors whose PDE is equal to (a) 6\% and (b) 20\%. Neutrinos are simulated along the Z axis. Tracks from both the muon and three stopping protons can be seen. The latter show a higher energy deposition due to the Bragg peak at the end of their tracks.
    \label{fig:Sim_numu_CCQE}
    }
\end{figure*}

Finally, to determine the number of photons produced at different positions along the fibres and detected at the SPAD array, the light yield was corrected for the attenuation length \cite{kuraray-catalogue-ps}. As discussed above, random noise is expected to be extremely small, in particular after applying the track algorithm selection. Thus, it is neglected in this study. An example of a $\nu_\mu$ CCQE event can be seen in Fig.~\ref{fig:Sim_numu_CCQE}. The proton Bragg peak, whose stopping power can be as high as 5--10 times that of a minimum ionizing particle (MIP), is well visible even for a relatively low PDE. This could make SPAD array sensors a viable solution for the detection of low-momentum protons already with existing technologies. 
Using the high PDE SPAD array sensor, the MIP can be tracked more precisely, even showing delta-electrons along the track.

\begin{figure*}
    \centering
    \begin{subfigure}{0.49\textwidth}    
            \includegraphics[width=\linewidth]{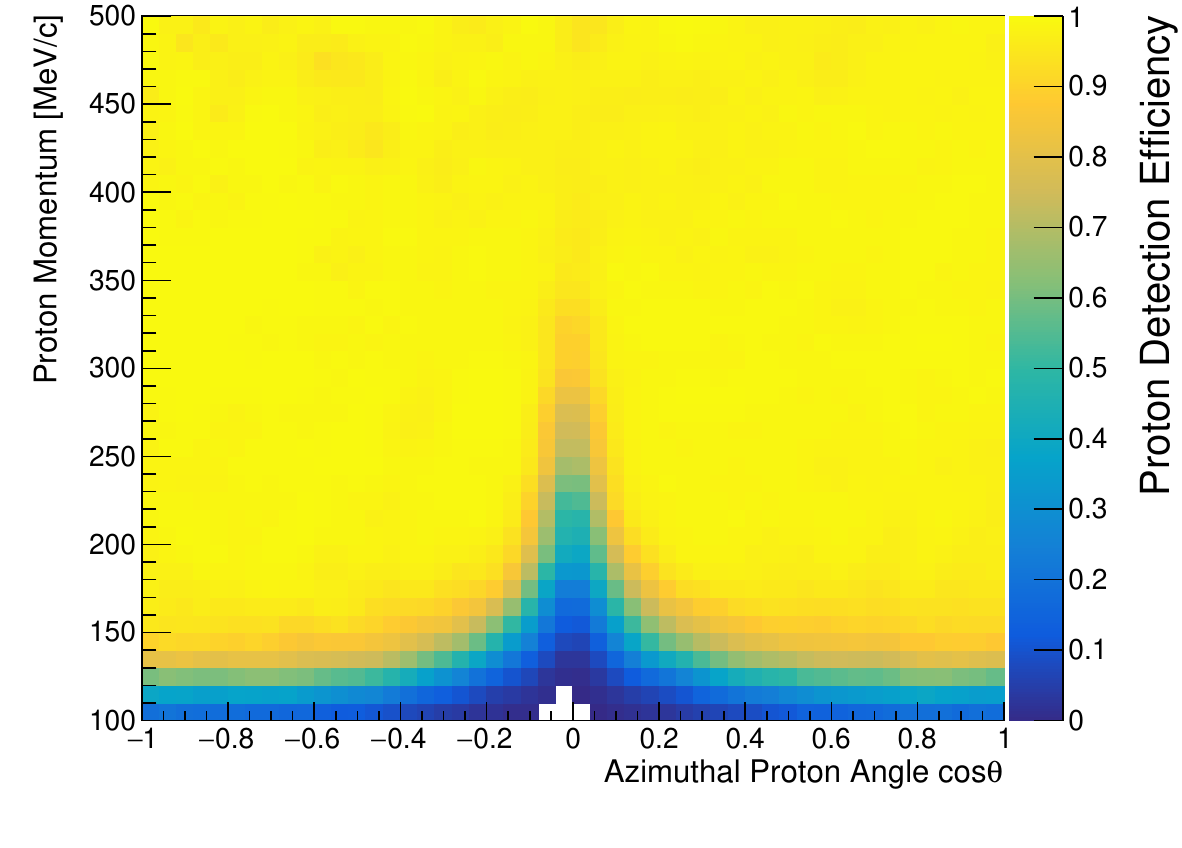} 
            \caption{}
            \label{fig:proton_efficency_SwissSPAD}
        \end{subfigure}
        \begin{subfigure}{0.49\textwidth}    
            \includegraphics[width=\linewidth]{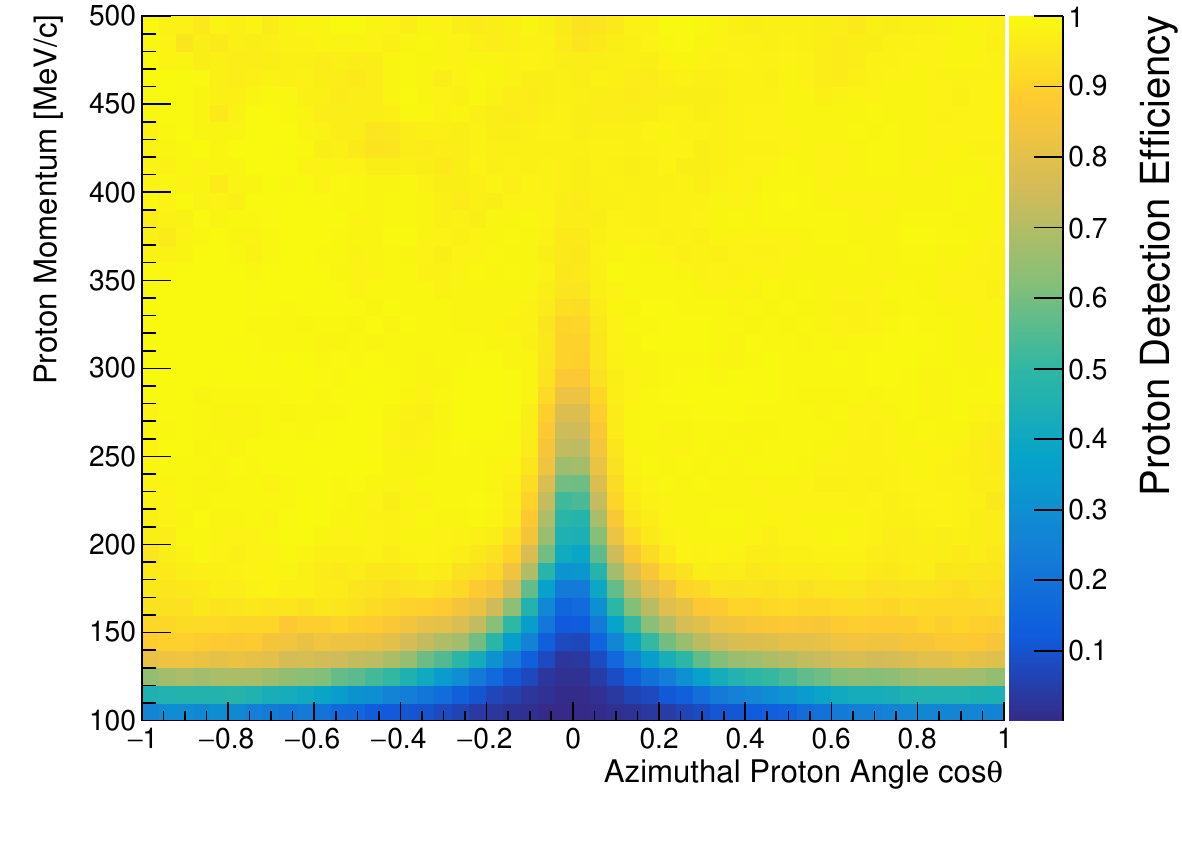} 
            \caption{}
            \label{fig:proton_efficency_PLATON}
        \end{subfigure}
        \begin{subfigure}{0.49\textwidth}    
            \includegraphics[width=\linewidth]{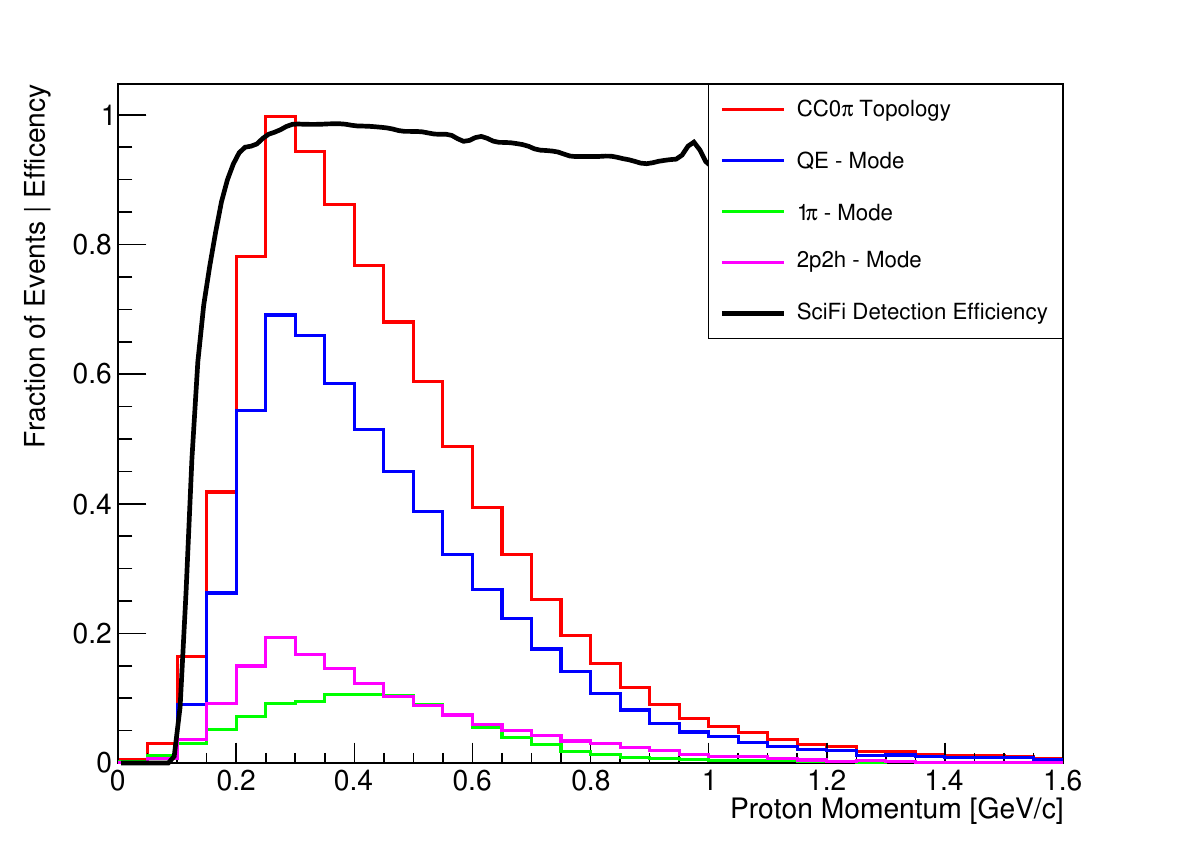} 
            \caption{}
            \label{fig:proton_Cutoff}
        \end{subfigure}   
    \caption{
    (a) Simulated proton tracking efficiency as a function of the proton momentum and direction (cosine of the angle with respect to the neutrino direction) with a 6\% PDE SPAD array, similar to the one used in the measurements in the previous sections.
    (b) Simulated proton tracking efficiency as a function of the proton momentum and direction (cosine of the angle with respect to the neutrino direction) with a 20\% PDE SPAD array, as proposed in this paper. 
    (c) Proton tracking efficiency overlaid to the momentum distribution of protons produced by $\nu_{\mu}$-carbon interactions without pions in the final state. The proton momentum distribution is broken down by the different neutrino interaction modes, i.e. CCQE, 2p2h and CC with one pion.
    The simulation was performed with NEUT~\cite{NEUTSim2021} using the T2K neutrino flux \cite{t2kfluxurl}.
    }
    \label{fig:proton_efficency}
\end{figure*}

A ``pseudo-reconstruction'' was implemented to study the physics potential of such a detector configuration. A ``hit'' is observed if at least one photon is detected by the SPAD array from a given fibre.
A track is reconstructed if it contains at least five hits, three in one projection and two in the orthogonal one. Moreover, in order to account for the overlap between different particle tracks in the tracking efficiency, the shorter track is assumed to be reconstructed if its last hit is separated by at least 2 fibres from the nearest track. The results of this study show that protons down to 150 MeV/c can be reliably detected, as shown in Fig.~\ref{fig:proton_efficency_PLATON}, thanks to the fine granularity of plastic scintillating fibres.
The notable drop in efficiency at $\cos\theta \sim 0$ is due to protons that, travelling along a single fibre, do not fulfill the reconstruction criteria.
In Fig.~\ref{fig:proton_Cutoff} the proton tracking efficiency is plotted as a function of the proton momentum and superimposed to the truth proton spectrum corresponding to $\nu_{\mu}$ CC interactions without pions in the final state.
A \scifi~detector gives access to the full spectrum of protons produced by neutrino-nucleus interactions and becomes a tool to resolve the nuclear effects and constrain the major systematics at the future LBL neutrino oscillation experiments.
It is interesting to note that, given the high stopping power, low-momentum protons can be efficiently reconstructed even with SPAD array sensors with a PDE down to about 6\%.
Studies investigating particle identification using the ionisation density along the path, charge identification and overall calorimetry are in progress.
However, a very good proton purity 
up to 95\% can be achieved
thanks to the Bragg peak which is much more pronounced compared to that of muons and pions.

Another promising study performed with the 20\% PDE configuration is the rejection of $\gamma \to e^+ e^-$ events, a major background to $\parbar{\nu}_e$ interactions.
Due to the high granularity of this type of detector, it is possible to resolve the initial tracks of both the electron and the positron, even in relatively low energy gamma conversion events.
In a 0.2 T magnetic field, it was found that for $\gamma \rightarrow e^+ e^-$ events between 0.1 and 0.5 GeV, the major background in the detection of $\nu_e$-nucleus interactions, only about 5\% cannot be fully reconstructed and properly identified. Such a fraction is largely dominated by Compton scattering where the observable is just a single electron track.  

\section{Conclusions}
\label{sec:conclusions}
We proposed a neutrino active target detector consisting of scintillating fibres read out with SPAD array sensors to detect and track particle events produced by neutrino interactions, such as low-momentum protons or electrons and positrons from photon conversion, key for the reduction of the systematic uncertainty at the future long-baseline neutrino oscillation experiments. 

SPAD array sensors are monolithic imaging devices where each SPAD pixel acts as an effectively independent read out channel, allowing to detect single photons with a spatial resolution below $10~\mu\text{m}$ 
with minimal impact from random noise and, depending on the chosen readout strategy, also the time of arrival with a time resolution below 100~ps.
One clear advantage is that the number of electronic channels scales with the detector surface rather than the number of scintillating fibres, with a potentially drastic reduction of the costs, despite an extremely fine detector granularity.

In this work we provided the first proof of concept of particle tracking with a state-of-the-art SPAD array sensor (SwissSPAD2) and analysed the main parameters to be optimised for future applications in neutrino experiments. 

The result of this work is being used to develop a new SPAD array sensor, on whose design the authors are currently working.
The current PDE is relatively low compared to SiPMs, due to the lower fill factor, and is the main parameter that needs to be improved, together with the single-pixel time resolution. 
On the other hand, it is worth noting that even a PDE below 10\% would allow to efficiently track protons below 300 MeV/c, thanks to the high-stopping power. Single-pixel time information would allow suppression of dark counts simply by coincidence, despite the size of the gate width. The authors are now also moving towards a more advanced prototyping of an XY \scifi~configuration, still read out with SPAD array sensors.

Simulation studies demonstrated the potential of \scifi~in the detection of particles produced by neutrino interactions, in particular those leaving very short tracks that are invisible to state-of-the-art tonne-scale neutrino detectors.

Finally, it is interesting to note that one may alternate \scifi~modules 
with thicker volumes of coarser 3D-segmented scintillator, like the one in \cite{Sgalaberna:2017khy}, to obtain a massive  detector where scintillating fibres can be used as neutrino active targets as well as for improving the reconstruction of particles produced by interactions in the coarser granularity part. Such a configuration could be optimal for combining high-resolution tracking with the efficient detection of neutrons that, as shown in \cite{Munteanu:2019llq,Gwon:2022bix}, require large volumes of dense hydrocarbon-like active material, or of electron neutrinos ($\nu_e$), that typically contaminate only a few percent of the $\nu_{\mu}$ dominated neutrino beam.
Moreover, like already done with nuclear emulsions \cite{NINJA:2020gbg,NINJA:2022zbi}, single layers of XY SciFi could be alternated with very thin volume of inactive target material to precisely measure neutrino interactions also in nuclei other than carbon or hydrogen, like water, but with the advantage of an excellent time resolution.

\section*{Statements and Declarations}
The authors declare that there are no conflicts of interest related to this article. For the sake of transparency, the authors would like to disclose that (i) Edoardo Charbon holds the position of Chief Scientific Officer of Fastree3D, a company making LiDARs for the automotive market, and that (ii) Claudio Bruschini and Edoardo Charbon are co-founders of Pi Imaging Technology. Both companies have not been involved with the paper drafting, and at the time of writing have no commercial interests related to this article.

\section*{Acknowledgments}
This work was supported by the SNF grant PCEFP2\textunderscore203261, Switzerland. This research was also partially supported by the Swiss National Science Foundation (grant 20QT21\textunderscore187716 Qu3D ``Quantum 3D Imaging at high speed and high resolution").

\bibliographystyle{unsrtnat}

\bibliography{bibliography}

\end{document}